\documentclass[journal]{new-aiaa} %for journal papers
\usepackage[utf8]{inputenc}
\usepackage{graphicx}
\usepackage{amsmath}
\usepackage[version=4]{mhchem}
\usepackage{siunitx}
\usepackage{longtable,tabularx}
\usepackage{subfigmat}
\usepackage{lipsum}                     % Dummytext
\usepackage{xargs} 
\usepackage[colorinlistoftodos,prependcaption,textsize=tiny]{todonotes}
\newcommandx{\unsure}[2][1=]{\todo[linecolor=red,backgroundcolor=red!25,bordercolor=red,#1]{#2}}
\newcommandx{\needswork}[2][1=]{\todo[linecolor=blue,backgroundcolor=blue!25,bordercolor=blue,#1]{#2}}
\newcommandx{\info}[2][1=]{\todo[linecolor=OliveGreen,backgroundcolor=OliveGreen!25,bordercolor=OliveGreen,#1]{#2}}
\newcommandx{\improvement}[2][1=]{\todo[linecolor=Plum,backgroundcolor=Plum!25,bordercolor=Plum,#1]{#2}}
\newcommandx{\thiswillnotshow}[2][1=]{\todo[disable,#1]{#2}}

\usepackage[normalem]{ulem}
\usepackage{color,soul}
\setulcolor{red}
\sethlcolor{yellow}
\title{Linear and Nonlinear Disturbance Evolution on the Frustum of Hypersonic Ogive-Cylinders}

\author{Hemanth Goparaju\footnote{Graduate Research Assistant, Mechanical and Aerospace Engineering}}
\affil{The Ohio State University, Columbus, Ohio, 43210}
\author{Kunal C. Kanawade\footnote{Graduate Research Assistant, Mechanical Engineering}}
\affil{Florida State University, Tallahassee, Florida, 32310}
\author{S. Unnikrishnan\footnote{Assistant Professor, Mechanical Engineering}}
\affil{Florida State University, Tallahassee, Florida, 32310}
\author{Datta V. Gaitonde\footnote{Glenn Chair Professor, Mechanical and Aerospace Engineering, AIAA Fellow}}
\affil{The Ohio State University, Columbus, Ohio, 43210}

\begin{document}
\maketitle
\begin{abstract}
  Aspects of transition mechanisms on a $14^{\circ}$ sharp-nosed ogive-cylinder at Mach~$6$ are elucidated by considering linear and nonlinear disturbance evolution of freestream stochastic and wave packet forcing at different locations downstream of the ogive-cylinder junction.
  The spectral response of the stochastic forcing displays favorable agreement with experimental observations from AFRL on which the configuration is based.
  Intermittent wave packets are generated from small-amplitude continuous freestream pressure forcing. 
  Linear wave packet evolution reveals that Mack modes are the dominant primary instabilities, followed by relatively weaker first-mode waves.
  The non-linear wavepacket displays a three-legged wall pressure perturbation signature, which is traced to spanwise curvature effects, and fundamental resonance is the dominant secondary instability mechanism.
  At higher amplitudes of excitation, weak signatures of sub-harmonic and oblique resonance phenomena are identified. 
Downstream placement of the small-amplitude disturbance diminishes receptivity of first-mode waves; however, with high-amplitude forcing, the location of actuation does not significantly vary the disturbance evolution signature or the secondary instability mechanisms.
 
 \end{abstract}

\section*{Nomenclature}

{\renewcommand\arraystretch{1.0}
\noindent\begin{longtable*}{@{}l @{\quad=\quad} l@{}}
$M$  & Mach number \\
$\mu$ & dynamic viscosity coefficient \\
$\gamma$ & specific heat ratio \\
$Pr$  & Prandtl number \\
$Re$ & Reynolds number \\
$f$ & frequency [kHz]\\
$\omega$ & angular frequency \\
$(\alpha, \beta)$ & streamwise, spanwise wavenumbers  \\
$\lambda_z$ & spanwise wavelength \\
$\delta_h$   & boundary-layer thickness \\
$\epsilon$ & amplitude parameter \\
$r$ & random number \\
$\rho$ & density     \\
$p$ & pressure \\
$T$ & temperature \\
$t$ & time coordinate\\
$e$ & total energy \\
$H$ & specific enthalpy \\
$Q$ & vectors of conserved variables in body-fitted coordinates\\
$(u, v, w)$& streamwise, wall-normal, spanwise velocity components in Cartesian coordinates\\
$(\xi, \eta, \zeta)$  & streamwise, wall-normal, azimuthal coordinates in body-fitted coordinate system \\
$(x, y, z)$  & streamwise, wall-normal, azimuthal coordinates in Cartesian coordinate system \\
$(E, G, F)$ & flux vectors in $(\xi, \eta , \zeta)$ directions \\
$J$ & transformation Jacobian matrix \\
$\tau$ & stress tensor\\
$\Bar{I}$ & heat flux \\
\multicolumn{2}{@{}l}{Subscripts}\\
$\infty$ & Freestream conditions\\
$e$ & boundary-layer edge \\ 
$w$ & wall conditions\\
$n$ & wall -normal component\\
$C$ & vectors in Cartesian coordinates \\
\multicolumn{2}{@{}l}{Superscript}\\
$'$ & perturbation \\
\end{longtable*}}

\section{Introduction}
The location of laminar-turbulent transition in high-speed boundary layers is crucial in the design of hypersonic aircraft because of its influence on drag and wall heating.
The development of accurate models to predict this location is a challenging endeavor because of the numerous possible transition paths, which depend on the nature of mean flow and freestream disturbances~\cite{morkovin1987transition}.
In wind tunnels, hypersonic transition experiments have mainly been performed on canonical configurations such as flat plates, wedges, and cones.
A comprehensive documentation of studies until the beginning of the 21st century may be found in Schneider~\cite{schneider2004hypersonic}.
Numerous recent ground test experiments have greatly enriched the available databases.
Mean flow variations have been incorporated by varying the leading edge bluntness, angle of attack, and surface roughness.
Mean flow modifications due to ogive forebodies are known to result in improved drag and heating performance. 
Experimental and numerical studies characterizing transition in these configurations are, however, relatively scarce.
The present work focuses on a numerical examination of  transition mechanisms on the frustum of a sharp ogive cylinder.
The configurational parameters are anchored in recent conventional (noisy) Air Force Research Laboratory Mach-6 Ludwieg Tube experiments by Hill et al.~\cite{luke2022experimental}, which are described in Sec.~\ref{sec:Base}.
%\todo{Removed noisy-tunnel from here -- will need to add it later HG: Good to clarify about noisy conditions in the introduction.}

Transition in convectively unstable boundary layers may be viewed as a forced response of a complex oscillator~\cite{huerre1990local}. 
The boundary-layer essentially acts as an amplifier, with its transition depending on the nature and spectral content of freestream disturbances, and their selective amplification~\cite{novikov2016direct}.
Freestream disturbances have low amplitudes in typical atmospheric flight scenarios, where transition to turbulence displays three main phases including, receptivity, linear disturbance amplification, and nonlinear modulation leading to turbulence.
Receptivity is the process through which environmental fluctuations penetrate the boundary layer.
In low-speed flows at zero angle of attack, the evolution of linear disturbances is described in terms of planar Tollmien-Schlichting waves~\cite{reshotko1976boundary}.
With increasing flow speeds to the supersonic regime, the most amplified waves are oblique first modes.

At even higher speeds, in hypersonic flows, a direct resonance is observed between freestream acoustic waves and boundary layer modes~\cite{fedorov2001prehistory} and the wavelength conversion mechanism of low-speed flows~\cite{saric2002boundary} is no longer needed. 
In sharp-nosed axisymmetric cones and flat plates, Mack~\cite{mack1984boundary} derived
%\todo{Gives impression Mack did work on axisymmetric cones as well.  True?HG:Yes.Mack did work on axisymmetric cones} 
the existence of a family of higher instability modes (called Mack modes).
At freestream Mach numbers higher than $4$, the first of these, designated the second-mode, displayed growth rates exceeding those of the first mode~\cite{masad1995relationship}.
The characterization of the second mode in terms of fast and slow acoustic waves and the $F$ and $S$ instability nomenclature has been discussed in detail by Fedorov and Tumin~\cite{fedorov2011high}. %\todo{Enter citation and any additional characterization that might be useful. We need this else the mode F below comes out of the blue. HEMANTH WILL DO. HG:Done}
Under typical wind tunnel conditions,
Kendall~\cite{kendall1975wind} established second-mode disturbances as the dominant primary instabilities for axisymmetric cones.
More recently, Scholten et al.~\cite{scholten2022linear} used modal stability analyses, which do not account for receptivity,  to identify Mack modes as the dominant primary instabilities for hypersonic sharp ogive cylinders.
% However, the method does not account for the receptivity of these instabilities.
In the last stage of the transition, when amplitudes are sufficiently high, the primary instabilities saturate and transfer energy to other disturbances through triadic interactions~\cite{herbert1988secondary}.
These disturbances interact non-linearly, leading to spectral filling and turbulence.

During transition, the nature and composition of freestream perturbations set the initial conditions (frequency, phase, amplitude) that dictate subsequent disturbance evolution. 
In wind tunnels and free flight, perturbation sources include acoustic radiation from tunnel walls, entropic spots, vortical gusts, and particulates~\cite{bushnell1990notes}.
It is difficult to numerically specify the composition of these disturbances, as they remain largely unknown~\cite{schneider2004hypersonic}.
However, various more recent efforts have been made to model the broadband nature of the tunnel or free-flight disturbances.
Balakumar and Chou~\cite{balakumar2018transition} simulated freestream disturbances based on Mach~$10$ sharp cone experiments by Marineau et al.~\cite{marineau2015investigation} to successfully predict the location of the onset of the transition.
Reduced order models to inform the inlet disturbances can be developed if a high-fidelity tunnel simulation is available.
For example, Goparaju et al.~\cite{goparaju2022supersonic}, Liu et al.~\cite{liu2022interaction} developed such models based on Duan et al.~\cite{duan2019characterization} turbulence data sets and captured first- and second-mode driven breakdown at Mach~$2.5$ and $8$, respectively. 

In the absence of relevant experimental or numerical wind tunnel data, freestream perturbations have been successfully modeled as stochastic perturbations following the work of Hader and Fasel~\cite{hader2018towards}, who used the approach to show that fundamental breakdown is the dominant route to turbulence in flared cones at Mach~$6$.
Later, Goparaju et al.~\cite{goparaju2021effects} used this perturbation specification approach to examine the role of leading edge bluntness in hypersonic flat plates.
The evolution of linear disturbances highlighted a change in the nature of instability waves at higher nose radii.
Since a detailed composition of freestream perturbations in the tunnels used by Hill et al.~\cite{luke2022experimental} is currently lacking, we use this stochastic approach to model inlet pressure perturbations in Sec.~\ref{sec:RandLin}.

The role of other sources of disturbances during transition described by Bushnell~\cite{bushnell1990notes} has also been quantified in the literature.
Fedorov et al.~\cite{fedorov2013temp} modeled freestream entropy-spots to note the generation of instability  mode~$F$~\cite{fedorov2011high} inside the boundary layer, and downstream excitation of mode~$S$ (second mode in this case) through the intermodal exchange mechanism.
Second-mode receptivity of slow freestream acoustic waves was shown to be stronger than that of fast acoustic waves on supersonic flat plates by Ma and Zhong~\cite{ma2005receptivity}.
Furthermore, freestream entropy and vorticity receptivity were very similar to those of freestream fast acoustic waves.
On hypersonic cylinder-wedges, Cerminara and Sandham~\cite{cerminara2017acoustic} showed strong resonant amplification of the $F$-mode closer to the leading edge with freestream fast acoustic forcing.

In addition to stochastic freestream excitation, we also consider wavepacket forcing, which has proven very insightful in isolating transition mechanisms in a controlled manner. 
The physical motivation for such excitation stems from the fact that solid particulate impact can excite instability waves in high speed boundary layers, as elucidated in the theoretical model of Fedorov~\cite{fedorov2013receptivity}. %predicts the excitation of instability waves in high speed boundary layers upon particle impact.
Direct numerical simulations of Chuvakhov et al~\cite{chuvakhov2019numerical} demonstrate that particle interactions can indeed generate wave packets; this  was shown to result in transition on a $14^{\circ}$ half-angle wedge at Mach~$4$.
%Fedorov and Obraz
Hader and Fasel~\cite{hader2021three} showed that the evolution of a localized pulse (wave packet) is a good surrogate to model particle impact effects.
These results have led to the extensive use of wave packet analysis to simulate aspects of natural transition, i.e., through broadband disturbance initiation (see, \cite{hader2021three,sivasubramanian2014numerical}).
In tests, such pulses were generated on hypersonic wind tunnel walls by Casper et al.~\cite{casper2014pressure} through glow discharges, and the evolution of disturbances was tracked from the linear to the breakdown stages. 

The goal of the present work is to examine transition mechanisms on ogive cylinders, so as to complement analogous studies on cones~\cite{sivasubramanian2014numerical, hader2021three} and flat plates~\cite{sivasubramanian2016reynolds, unnikrishnan2020linear}.
%Although nonlinear transition stages have been well
%there is no corresponding study for ogive cylinders in the literature.
The key consideration that distinguishes ogive cylinders from these configurations is the presence of significant spanwise curvature throughout the streamwise length.
%, which are absent in the above configurations. 
The linear and nonlinear stages of transition in these geometries are addressed in Sec.~\ref{sec:WP} through wave packet analysis.  

This paper is organized as follows.
The geometry and numerical methods are discussed in Sec.~\ref{sec:flow}.
Steady base flow features, followed by linear and nonlinear disturbance evolution, are examined in Sec.~\ref{sec:results}.
The conclusions of this work are provided in Sec.~\ref{sec:conclusions}.

\section{Flow Configuration and Numerical Methodology}\label{sec:flow}
The forebody geometry and the freestream parameters are first described, followed by a description of the numerical methods and procedures used.
% to examine the evolution of the disturbances.
\subsection{Geometry}\label{sec:geometry}
\begin{figure}
    \centering
    \includegraphics[width=0.8\textwidth]{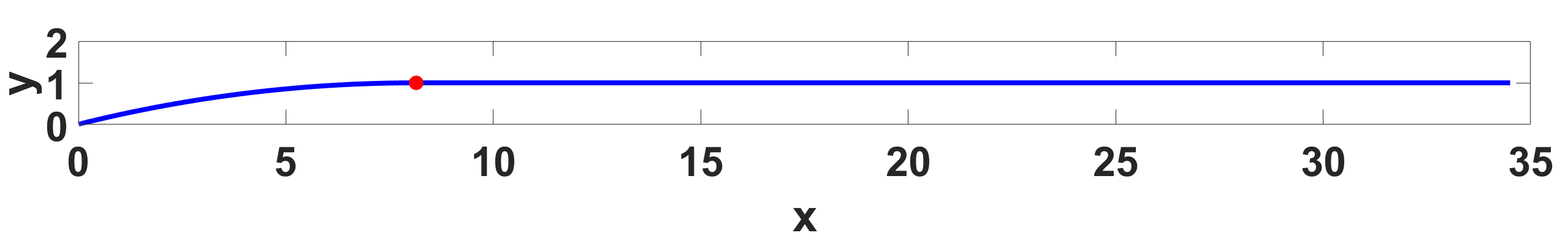}
    \caption{Profile of the ogive-cylinder forebody with junction marked in red.}
    \label{fig:profile}
\end{figure}

The forebody mimics the Mach~6 experiments of Hill et al.~\cite{luke2022experimental} and is comprised of an ogive cylinder of half-angle $14^{\circ}$ and nose radius ~$0.1 mm$, which is considered to be sharp. 
The forebody length is $0.253m$ beyond which it is tangentially joined to a cylinder of radius~$0.0311 m$, and length~$1 m$.
Lengths are non-dimensionalized by the cylinder radius; the body profile in
terms of the streamwise ($x$) and wall-normal ($y$) directions is shown in Fig.~\ref{fig:profile}.
%\todo{should it be y or r for wall normal? FIXED.} 
The ogive-cylinder junction is thus located at x=$8.135$ (marked in the figure) and the cylinder section extends until the end of the domain. 

%???it is better to provide a geometry figure here with the dimensions. kunal, you should be able to this very easily. Done 
%???menion the x-value of the ogive-cylinder junction. Done

The freestream parameters are based on the Mach~6 Ludwieg Tube wind tunnel conditions at the Air Force Research Laboratory (AFRL).
The angle of attack is specified to be zero, and the freestream Mach number is $6.1$.
The oncoming flow is at a Reynolds number~$Re=7.01\times 10^{6} /m$.
The temperature of the freestream $(T_{\infty})$ is $58.01 K$, and the wall is isothermal, $(T_{w})$ at $300K$.
This results in a slightly cooled wall relative to the recovery temperature for these freestream conditions.

\subsection{Numerical Techniques and Boundary Conditions} \label{sec:numerical}
The flow is modeled using unsteady viscous compressible Navier-Stokes equations, cast in the strong conservative form as:
\begin{equation}
\frac{\partial \boldsymbol{Q}}{\partial t}+\frac{\partial \boldsymbol{E}}{\partial \xi}+\frac{\partial \boldsymbol{G}}{\partial \eta}+\frac{\partial \boldsymbol{F}}{\partial \zeta}=0
\end{equation}
where $t$ is the time and $(\xi, \eta, \zeta)$ are the spatial coordinates.
$\textbf{Q}$ is the vector of conservative variables, and $(\textbf{E},\textbf{G},\textbf{F})$ are flux vectors that satisfy the relation:
\begin{equation}
\begin{gathered}
\boldsymbol{Q}=J \boldsymbol{Q}_{c}, \quad \boldsymbol{E}=J\left(\boldsymbol{E}_{c} \frac{\partial \xi}{\partial x}+\boldsymbol{G}_{c} \frac{\partial \xi}{\partial y}+\boldsymbol{F}_{c} \frac{\partial \xi}{\partial z}\right) \\
\boldsymbol{G}=J\left(\boldsymbol{E}_{c} \frac{\partial \eta}{\partial x}+\boldsymbol{G}_{c} \frac{\partial \eta}{\partial y}+\boldsymbol{F}_{c} \frac{\partial \eta}{\partial z}\right), \\
\boldsymbol{F}=J\left(\boldsymbol{E}_{c} \frac{\partial \zeta}{\partial x}+\boldsymbol{G}_{c} \frac{\partial \zeta}{\partial y}+\boldsymbol{F}_{c} \frac{\partial \zeta}{\partial z}\right)
\end{gathered}
\end{equation}
Here,  $\textbf{J}=det[\partial (x,y,z)/\partial (\xi, \eta, \zeta)]$ is the Jacobian of the coordinate transformation and  $(\textbf{E}_c,\textbf{G}_c,\textbf{F}_c)$ are given by:
\begin{equation}
\begin{aligned}
&{\boldsymbol{Q}_{c}=\left[\begin{array}{c}\rho \\ \rho u \\ \rho v \\ \rho w \\ \rho e\end{array}\right], \quad \boldsymbol{E}_{c}=\left[\begin{array}{c}\rho u \\ \rho u u+p+\tau_{x x} / R e_{\infty} \\ \rho u v+\tau_{x y} / R e_{\infty} \\ \rho u w+\tau_{x z} / R e_{\infty} \\ \rho u H+I_{x} / R e_{\infty}\end{array}\right]}
&{\boldsymbol{G}_{c}=\left[\begin{array}{c}\rho v \\ \rho v u+\tau_{y x} / R e_{\infty} \\ \rho v v+p+\tau_{y y} / R e_{\infty} \\ \rho v w+\tau_{y z} / R e_{\infty} \\ \rho v H+I_{y} / R e_{\infty}\end{array}\right], \quad \boldsymbol{F}_{c}=\left[\begin{array}{c}\rho w \\ \rho w u+\tau_{z x} / R e_{\infty} \\ \rho w v+\tau_{z y} / \operatorname{Re}_{\infty} \\ \rho w w+p+\tau_{z z} / R e_{\infty} \\ \rho w H+I_{z} / \operatorname{Re}_{\infty}\end{array}\right]}
\end{aligned}
\end{equation}
The total energy per unit volume~$(e)$ and the total specific enthalpy~$(H)$ are obtained from:
\begin{equation}
   \begin{aligned}
e &=\frac{p}{\rho(\gamma-1)}+\frac{1}{2}\left(u^{2}+v^{2}+w^{2}\right), \\
H &=\frac{T}{\gamma-1 M_{\infty}^{2}}+\frac{1}{2}\left(u^{2}+v^{2}+w^{2}\right)
\end{aligned} 
\end{equation}
The components of stress tensor~$(\tau)$ are: 
\begin{equation}
\begin{gathered}
\tau_{x x}=-\mu\left(2 \frac{\partial u}{\partial x}-\frac{2}{3} \operatorname{div} V\right), \quad \tau_{y y}=-\mu\left(2 \frac{\partial v}{\partial y}-\frac{2}{3} \operatorname{div} V\right) \\
\tau_{z z}=-\mu\left(2 \frac{\partial w}{\partial z}-\frac{2}{3} \operatorname{div} V\right) \\
\tau_{x y}=-\mu\left(\frac{\partial u}{\partial y}+\frac{\partial v}{\partial x}\right), \quad \tau_{x z}=-\mu\left(\frac{\partial u}{\partial z}+\frac{\partial w}{\partial x}\right) \\
\tau_{y z}=-\mu\left(\frac{\partial v}{\partial z}+\frac{\partial w}{\partial y}\right)
\end{gathered}
\end{equation}
The heat flux vector components are:
\begin{equation}
\begin{gathered}
I_{x}=\left(u \tau_{x x}+v \tau_{x y}+w \tau_{x z}\right)-\lambda \frac{\partial T}{\partial x} \\
I_{y}=\left(u \tau_{y x}+v \tau_{y y}+w \tau_{y z}\right)-\lambda \frac{\partial T}{\partial y} \\
I_{z}=\left(u \tau_{z x}+v \tau_{z y}+w \tau_{z z}\right)-\lambda \frac{\partial T}{\partial z} \\
\lambda=\frac{\mu}{\operatorname{Pr}(\gamma-1) M_{\infty}^{2}}
\end{gathered}
\end{equation}
The equation of state closes the equations:
\begin{equation}
p=\frac{\rho T}{\gamma M_{\infty}^{2}}
\end{equation}

As noted earlier length-scales are normalized by the radius of cylindrical section ($l=0.0311~m$). 
Velocity, density, and temperature are normalized by the corresponding freestream quantities, while pressure is normalized by twice the freestream dynamic pressure~$(\rho_{\infty}u_{\infty}^2)$.  
Unless specified otherwise, all quantities are reported in nondimensional form.
%\todo{do we ever show dimensional quantities below? FIXED}
The working gas is assumed to be calorically perfect with Prandtl number~$(Pr=0.72)$, and ratio of specific heat~$\gamma=1.4$.
Viscosity is calculated using Sutherland's law, $\mu = T^{3/2}(S+1)/(S+T)$, where $S=110/T_{\infty}$, and the second viscosity is modeled by the Stokes hypothesis. 

The unsteady $3D$ simulations are predicated on precursor base flow simulations that include the leading edge of the geometry are performed using an axisymmetric solver formulated in the generalized cylindrical coordinates\cite{sandberg2007governing}. 
The grid has $3{,}251\times 601$ points in the streamwise~$(x)$ and wall-normal~$(y)$ directions, respectively.
%\todo{wasn't y the body normal direction? HG- yes. changed it to wall-normal}
$50\%$~of wall-normal points are clustered in the boundary-layer, and in the streamwise extent, the wave corresponding to largest frequency is resolved by $40$~points.
These are comparable to earlier transitional numerical simulations (see, {\em e.g.}, Ref. ~\cite{novikov2016direct}).
%\todo{do we have a citation that says this level of spacing is ok? HG: FIXED.}
Since the precursor calculations are designed primarily to capture the laminar basic state and shock, second-order accurate schemes are used for the convective and viscous fluxes with artificial dissipation for numerical stability\cite{jameson1981numerical}. 
%\todo{OK for me to write this?HG-yes}
Symmetry conditions are imposed along the axis upstream of the nose, Dirichlet boundary conditions are imposed on the freestream boundary, the wall is treated as a no-slip isothermal boundary with $T_w/T_{\infty}=5.16$, and 
a zero-gradient condition is applied on the downstream outflow boundary.

After the convergence of the axisymmetric base flow, the domain downstream of the ogive-cylinder junction is extracted from $x=9.0$, and rotated in the azimuthal direction within the range $\theta \in [-90^{\circ}, 90^{\circ}]$ for three-dimensional calculations using~$225$ azimuthal points.
The base flow is reconverged in time with the higher-order numerical scheme described below to serve as the basic state for perturbation evolution.
%Primary instabilities are resolved with~$40$ points per wavelength, which is comparable to previous transition studies~\cite{novikov2016direct}. 
Time-marching is performed using the second-order implicit Beam and Warming technique~\cite{beam1978implicit}.
Inviscid fluxes are reconstructed with the seventh-order WENO scheme~\cite{balsara2000monotonicity}, and solved with the Roe-Riemann solver~\cite{roe1981approximate}.
Viscous fluxes are evaluated through sixth-order compact-like schemes~\cite{visbal2002use}.
At the inlet, the first five streamwise points from the $2D$ computations are extracted from the wall-normal data.
These values are converted to cylindrical coordinates and imposed on the azimuthal plane as Dirichlet boundary conditions for the $3D$ computations.
The wall and freestream boundary conditions remain the same.
At the azimuthal boundaries i.e., the azimuthal velocity~$(v_{\theta})$ is antisymmetric while the radial velocity~$(v_r)$ is symmetric.
%\todo{I'm not sure I understand this radial/azimuthal thing HG: radial and azimuthal are used interchangeably. Should I stick to single convention?}
Zero-gradient conditions are imposed at the exit of the domain.
%\todo{spanwise may be better written as azimuthal. also asymmetric or antisymmetric? HG: antisymmetric sounds more clear. Changed it.}

\subsection{Disturbance Generation}
\label{sec:forcing}
The characteristics of disturbance evolution on the converged base flows are examined separately with stochastic and wave packet forcing. 
For stochastic forcing, pressure perturbations~$(p')$ are specified on the inlet plane:
%(after Dirichlet boundary condition) as:
\begin{equation}
    \frac{p'}{\rho_{\infty} u_{\infty}^2}=\epsilon (2r-1),
\end{equation}
where $\epsilon$ is the amplitude of the forcing.
Random numbers~$r\in [0,1]$ are generated with an inbuilt FORTRAN function following the work of Hader and Fasel~\cite{hader2018towards}.
This generates spatio-temporally varying disturbances at the inlet plane.
For two-dimensional studies, this forcing results in the evolution of axisymmetric disturbances while for the three-dimensional cases, the forcing has a delta form in the azimuthal direction, generating spanwise gradients.
%\todo{is this correct? HG:Yes.}.

Wavepacket forcing imposes localized pulses with a forcing function centered around the frequency~$\omega$:
\begin{equation}
\label{eq:WBS}
\begin{aligned}
&q_{w}(x, z, t)=(\rho v_{n})_{w}=\varepsilon \sin ^{3} \left(2 \pi \frac{x-x_{1}}{x_{2}-x_{1}}\right) \sin^{3} \left(\pi \frac{z-z_{1}}{z_{2}-z_{1}}\right) \sin \left(\omega t\right), \\
&x_{1} \leq x \leq x_{2}, \quad z_{1} \leq z \leq z_{2}, \quad 0 \leq t \leq t_{1}
\end{aligned}
\end{equation}
The wall-normal momentum is excited,  ensuring smooth gradients at the actuator edges,
by blowing-suction in the streamwise extent $x\in [x_1, x_2]$ as a dipole of width~$0.02$. 
In unrolled azimuthal coordinates, the slot $z\in [z_1, z_2]$ is excited by a monopole forcing function with width~$0.01$.
A wave packet at frequency $\omega$ is triggered in $t\in [0, t_1]$, with $t_1=2\pi/\omega$.
The amplitude~$\epsilon$ can be varied to generate waves of different strengths.
Various cases are examined in this work, encompassing linear and nonlinear receptivity and disturbance evolution, and are summarized in Table~\ref{tab:cases}.
\begin{table}
\caption{Simulation cases considered in this work} % title of Table
\centering 
\begin{tabular}{c c c c } 
\hline\hline %inserts double horizontal lines
Case & Forcing Nature & Forcing Location & Amplitude \\ 
\hline 
$1$ & Stochastic forcing- 2D & $x=9.0$ & Linear \\
$2$ & Stochastic forcing- 3D & $x=9.0$ & Linear \\
$3$ & Wave packet- 3D & $x=9.46$ & Linear \\
$4$ & Wave packet- 3D & $x=16.2$ & Linear \\
$5$ & Wave packet- 3D & $x=9.46$ & Nonlinear \\
$6$ & Wave packet- 3D & $x=16.2$ & Nonlinear \\
\hline\hline %inserts single line
\end{tabular}
\label{tab:cases} % is used to refer this table in the text
\end{table}

\section{Results}
% In this Section, the features of steady base flow are first discussed.
% It is followed by an examination of the response of the boundary layer to linear stochastic forcing.
% The evolution of linear and nonlinear wave packets is then elucidated from the receptivity to the secondary instability stages of transition.  
\label{sec:results}
\subsection{Steady Base Flow}
\label{sec:Base}
\begin{figure}
\centering
\begin{subfigmatrix}{1}
\subfigure[]{{\includegraphics[trim=150 20 60 350, clip, width=0.75\textwidth]{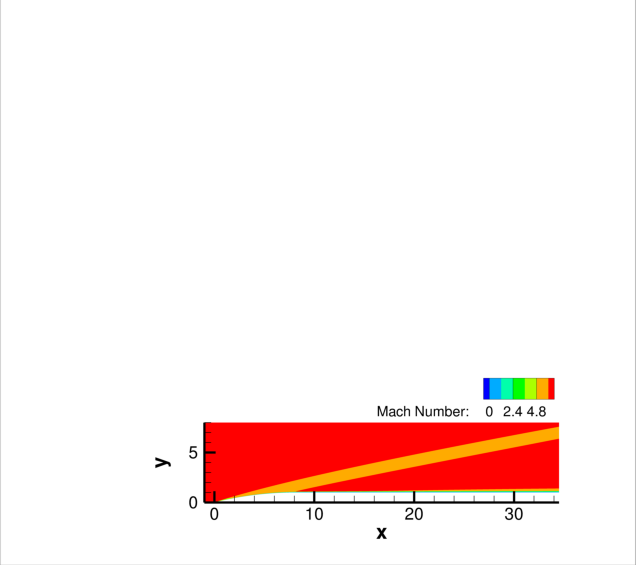}}}
\subfigure[]{{\includegraphics[trim=0 70 0 80, clip, width=0.75\textwidth]{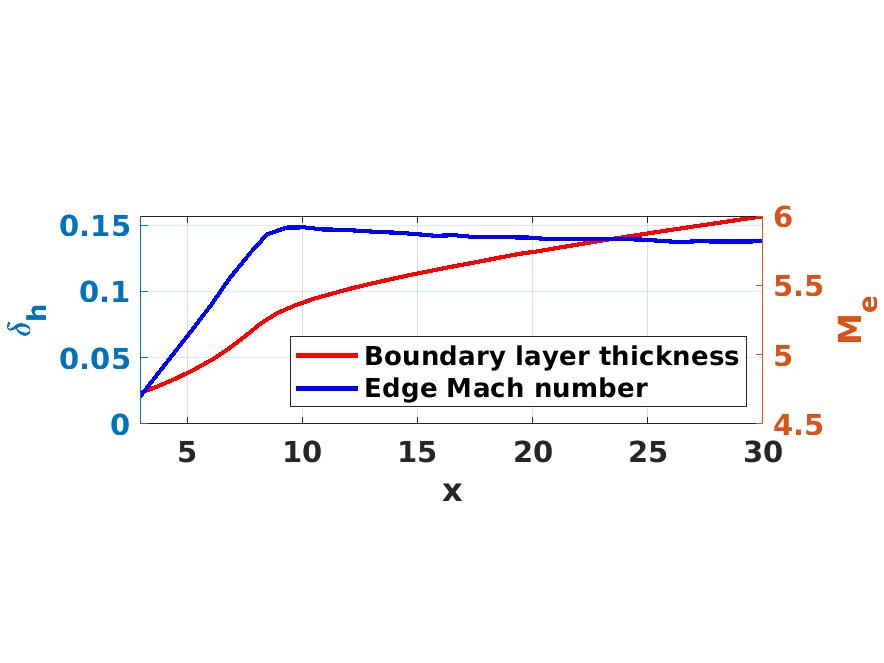}}}
%\subfigure[]{{\includegraphics[trim=0 0 0 0, clip, width=0.55\textwidth]{Ogv14_embl.png}}}
\end{subfigmatrix}
\caption{Steady base flow features of ogive-cylinder depicting
$(a)$ Mach number contours
$(b)$ Boundary layer edge properties.}
\label{fig:Ogv14_mach}
\end{figure}

An accurate prediction of the steady base flow is vital for the study of small-amplitude perturbation evolution~\cite{novikov2016direct}.%\todo{Is there a classical citation for this which shows what bad things happen if the basic state is bad? HEMANTH WILL FIX HG:Done}
The steady base flow features for the ogive cylinder are shown in Fig.~\ref{fig:Ogv14_mach}.
The Mach number contours in Fig.~\ref{fig:Ogv14_mach}(a) show the shock structure.
The detached shock region is associated with the relatively small ogive nose radius; it subsequently curves because of favorable pressure gradients arising from the curved ogive portion, and ultimately weakens to a Mach wave at the exit of the domain. 
A comparison of velocity and entropy profiles revealed that no significant entropy gradients exist outside the boundary layer, due to the relatively small shock curvature radius on the ogive portion.
%No significant entropy gradients are apparent due to the relatively small shock curvature radius on the ogive portion.\todo{Can we tell entropy gradients from what is plotted or should we write ``not shown''or some such. UNNI fix this.}

The edge of the boundary layer $(\delta_h)$ is chosen as the wall normal distance where the ratio of total enthalpy to the freestream enthalpy reaches~$0.995$~\cite{gupta1990hypersonic}.
%???include a reference if you have, that uses the above criteria
The streamwise evolution of $\delta_h$,  shown in Fig.~\ref{fig:Ogv14_mach}(b), indicates a monotonic increase with a change in growth rate evident at the junction of the ogive with the cylinder.
The boundary layer edge Mach number~$(M_e)$, considered at a wall normal height $1.5\delta_h$, is also shown in Fig.~\ref{fig:Ogv14_mach}(b).
The value of about $M_e\sim 6.0$ over the length of the cylinder provides a preliminary indication of  the potential dominance of the Mack (or second) mode.
%\todo{please check my change in terminology Mack/second HG:It is okay}

\begin{figure}
    \centering
    \includegraphics[trim=0 0 0 0, clip, width=0.75\textwidth]{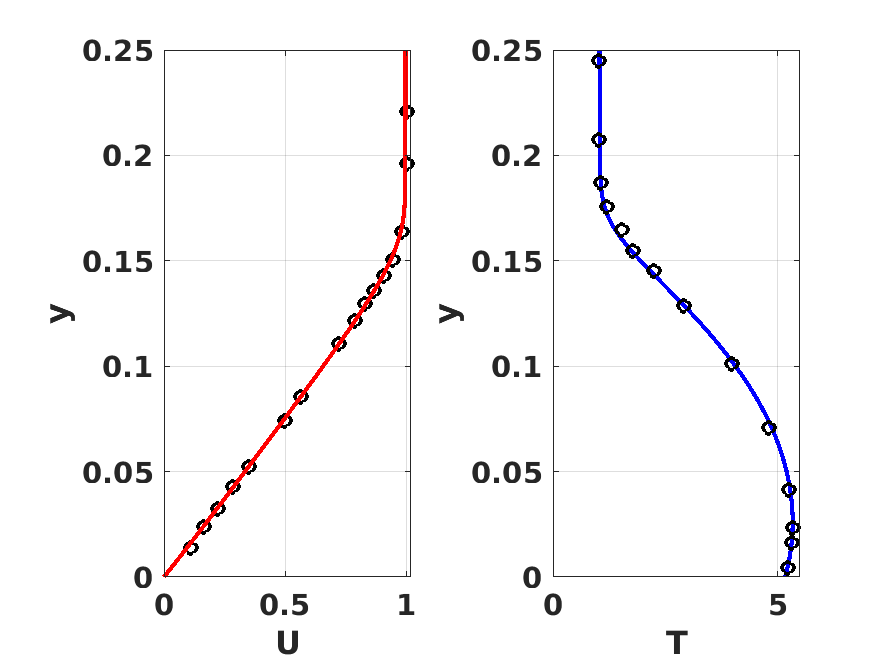}
    \caption{Wall-normal base flow profiles at $x=27.4$ for streamwise velocity~$(U)$ and temperature~$(T)$.
    The black circles~$(\circ)$ denote the profiles extracted from Scholten et al.~\cite{scholten2022linear}.}
    \label{fig:Valid}
\end{figure}

The wall-normal streamwise velocity~$(U)$ and temperature~$(T)$ profiles extracted near the end of the domain ($x=27$) or equivalently~$0.6m$ from the junction of the ogive-cylinder are shown in Fig.~\ref{fig:Valid}.
A good match with the corresponding profiles extracted from Scholten et al.~\cite{scholten2022linear} validates the accuracy of the current base flow.
%In the next Sections, disturbance evolution on this base flow is examined.

\subsection{Stochastic Linear Disturbance Evolution}
\label{sec:RandLin}
\subsubsection{Two-dimensional Perturbation Evolution}
\begin{figure}
\centering
{\includegraphics[trim=0 0 0 0, clip, width=0.75\textwidth]{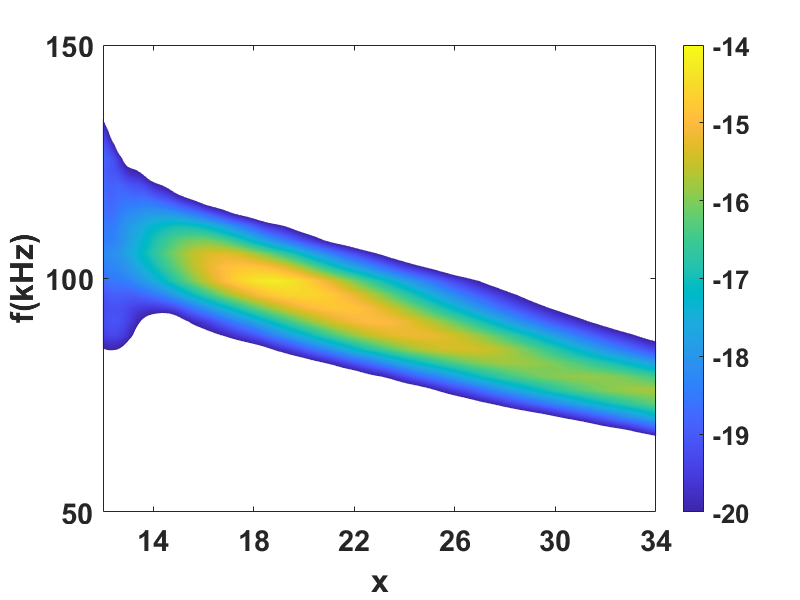}}
\caption{Frequency$(f)$- streamwise $(x)$ contours depicting the power spectral density of wall pressure perturbations with $2D$ stochastic forcing (Case~$1$).}
\label{fig:Ogv14_stx}
\end{figure}

The response of the axisymmetric boundary layer to  stochastic forcing from the freestream is first examined to characterize linear disturbance amplification, for which a small  amplitude of $\epsilon=1\times10^{-4}$~(Case~$1$) is used.
The power spectral density of the wall pressure perturbations versus the streamwise distance is displayed in Figure~\ref{fig:Ogv14_stx} .
%\todo{what is the quantity plotted? PSD? May help to clarify. It is a PSD plot - KK}
The most amplified disturbances lie in the relatively high-frequency range $(f\in [60-120kHz])$, and scale inversely with the thickness of the boundary layer, which are consistent with the properties of second mode instability~\cite{marineau2015investigation}.
%???kunal, re-scale the vertical axis range from 50 to 150. there is a lot of white space in the current figure. Done

\begin{figure}
\centering
\includegraphics[trim=0 0 0 0, clip, width=0.75\textwidth]{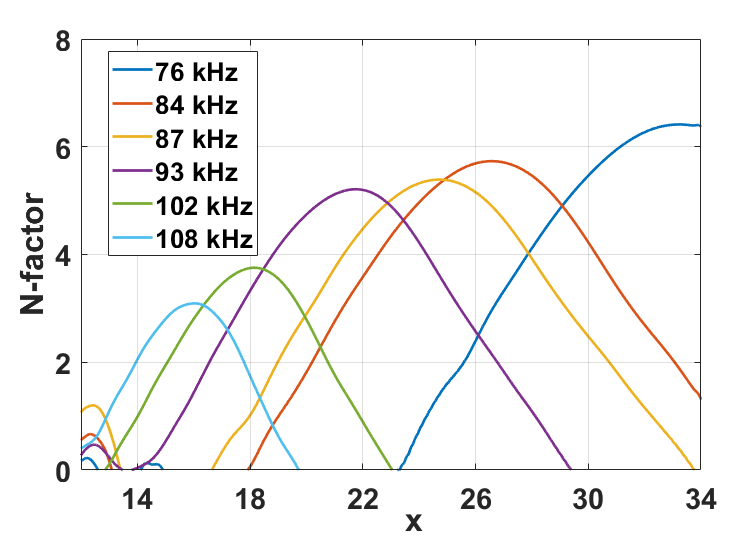}
\caption{N-factor curves from wall pressure perturbations comparing relative amplification, obtained from $2D$ stochastic forcing (Case~$1$).}
\label{fig:Ogv14_Nfac}
\end{figure}

The streamwise amplification of the disturbances is further quantified by the N-factor,  defined as the logarithm of the ratio of the amplitude of a perturbation quantity at a given $x$ location to its amplitude at a reference location, $x_0$. 
Here, the reference location is chosen as that where the linear growth in log scale is initiated in wall pressure perturbation data.
The N-factor curves, shown in Fig.~\ref{fig:Ogv14_Nfac}, indicate that higher frequencies are more amplified closer to the actuation location i.e., the lower frequencies indicate peaks at larger streamwise distances.
These curves are consistent with the modal instability analysis performed by Scholten et al.~\cite{scholten2022hypersonic}. 
The streamwise extent of the disturbance amplification zone increases with decreasing frequencies.
%\todo{Is this evident from this plot? If this is a forward reference, then that should be so indicated.HG:Fixed. It should be streamwise not spanwise} 
%???include a sentence like this - these n-factors are consistent with predictions in the modal analysis paper. Done.

\begin{figure}
    \centering
    \includegraphics[trim=0 0 0 0, clip, width=0.75\textwidth]{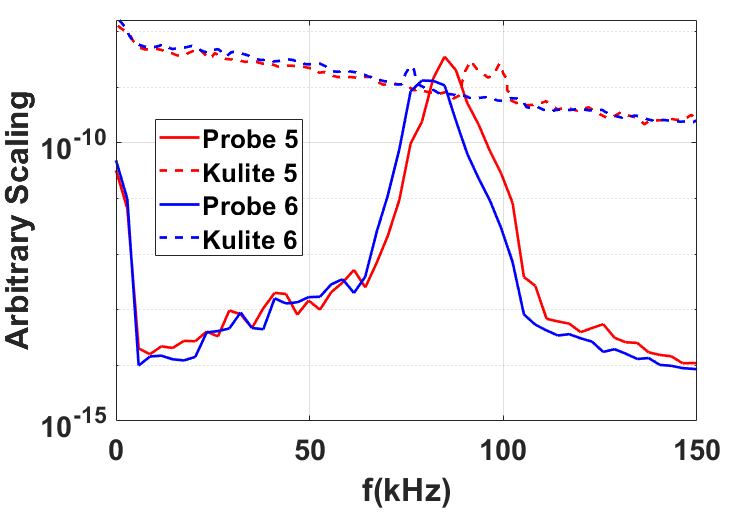}
    \caption{Comparison of experimental and numerical wall pressure spectral data (Case~$1$).
    The Kulite data is extracted from Hill et al.~\cite{luke2022experimental}.}
    \label{fig:Ogv14_psd1}
\end{figure}

The response to stochastic forcing is validated with the pressure transducer probes of Hill et al.~\cite{luke2022experimental} in Fig.~\ref{fig:Ogv14_psd1}.
The experimental measurements (dashed lines) and the corresponding computational probe measurements (solid lines) are compared for Kulites designated~$\#5$ and $\#6$, located at $x=26.7$ and~$29.6$ respectively, in the computational domain.
The peak amplification at these two locations correspond  to $f\sim85$ and~$75kHz$, respectively, which is in good agreement with the experimental observations.
This confirms that the numerical technique, including the chosen stochastic forcing, effectively captures the most amplified waves.  

\subsubsection{Three-dimensional Perturbation Evolution}
\begin{figure}
\centering
\begin{subfigmatrix}{2}
\subfigure[]{{\includegraphics[trim=5 1 30 15, clip, width=0.48\textwidth]{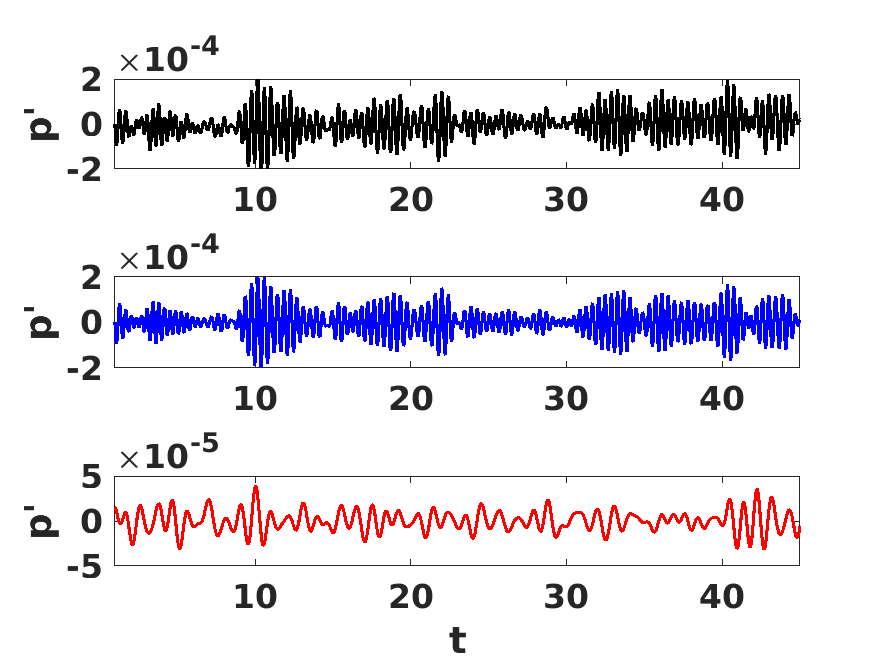}}}
\subfigure[]{{\includegraphics[trim=0 0 10 0, clip, width=0.48\textwidth]{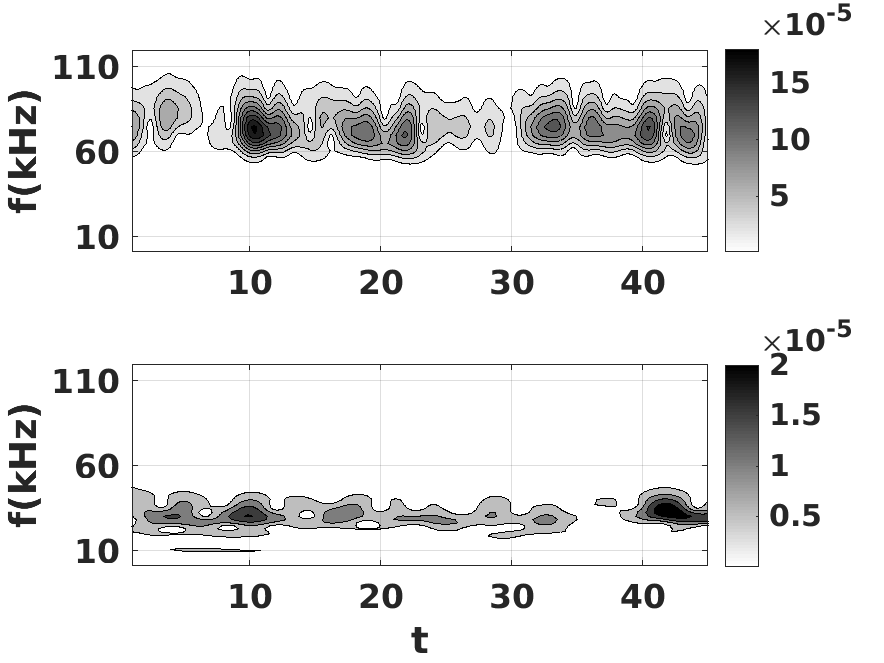}}}
\end{subfigmatrix}
\caption{Features of wall pressure perturbation probe at $x=25$ (Case~$2$).
In $(a)$ raw signal (top), bandpass signals $[60-150kHz]$ (middle), $[10-40kHz]$ (bottom) are shown.
In $(b)$ scalogram of high (top) and low (bottom) bandpass signals are depicted.}
\label{fig:DMD_Scalo}
\end{figure}

The three-dimensional response of the ogive cylinder boundary layer to broadband stochastic pressure fluctuations is examined using Case~$2$. 
To ensure linearity, an amplitude of $\epsilon=1\times10^{-3}$ is used.
Flow statistics are collected after the initial transients exit the domain.
The temporal features of the wall pressure perturbation at a typical streamwise location $(x=25)$ are shown in Fig.~\ref{fig:DMD_Scalo}. 
The disturbance signature (Fig.~\ref{fig:DMD_Scalo}(a), top) is modulated in amplitude, indicating that  despite the stochastic nature of the input, the perturbations organize and travel in groups as packets.
%\todo{There's use of a word ``passband'' in the caption of FIg. 7.  Is this standard usage? I've heard of bandpass as in high bandpass, but not passband. HG:Changed to bandpass}
%\todo{I changed this sentence. Please check. HG:OK}
This behavior is consistent with experimental~\cite{kennedy2019visualizations} and numerical stochastic transition studies~\cite{hader2018towards}. 

The wall pressure signal is filtered into frequency bands based on the response shown earlier in Fig.~\ref{fig:Ogv14_stx} of the $2D$ stochastic analysis.
The signal extracted in the range $f\in [60-150kHz]$ in Fig.~\ref{fig:DMD_Scalo}(a) (middle) exhibits similar structure as the original signal but with further amplitude modulation.
The corresponding frequency evolution (scalogram) in Fig.~\ref{fig:DMD_Scalo}(b) (top) attributes this to intermittent wave packets following each other.
These waves have a relatively broad spectral content in the second mode (frequency) range with higher amplitudes in the local (maximum) Mack mode wave at $f\sim 70kHz$.
The lower frequency content of the pressure signal extracted in the range $f\in[10-40kHz]$ (Fig.~\ref{fig:DMD_Scalo}(a), bottom) displays weaker amplitudes.
These disturbances have the characteristics of oblique first-mode waves, as they are not observed in Case~$1$.
The corresponding scalogram (Fig.~\ref{fig:DMD_Scalo}(b), bottom) also shows the wave intermittency behavior with amplification around $f\sim30kHz$. 

The evolution of disturbances is further characterized by identifying amplifying structures from the unsteady flow field.
Since structures at specific frequencies are desired, an optimal data-driven choice is the dynamic mode decomposition (DMD) algorithm of Schmid~\cite{schmid2010dynamic}.
The information obtained with this approach is similar to the Fourier transformation, where DMD provides a best-fit linear approximation with fewer snapshots~\cite{chen2012variants}.
Flow snapshots corresponding to the disturbance field,  extracted by subtracting the mean of the snapshots, are processed for both wall and symmetry planes.

\begin{figure}
\centering
{{\includegraphics[trim=0 0 0 0, clip, width=0.75\textwidth]{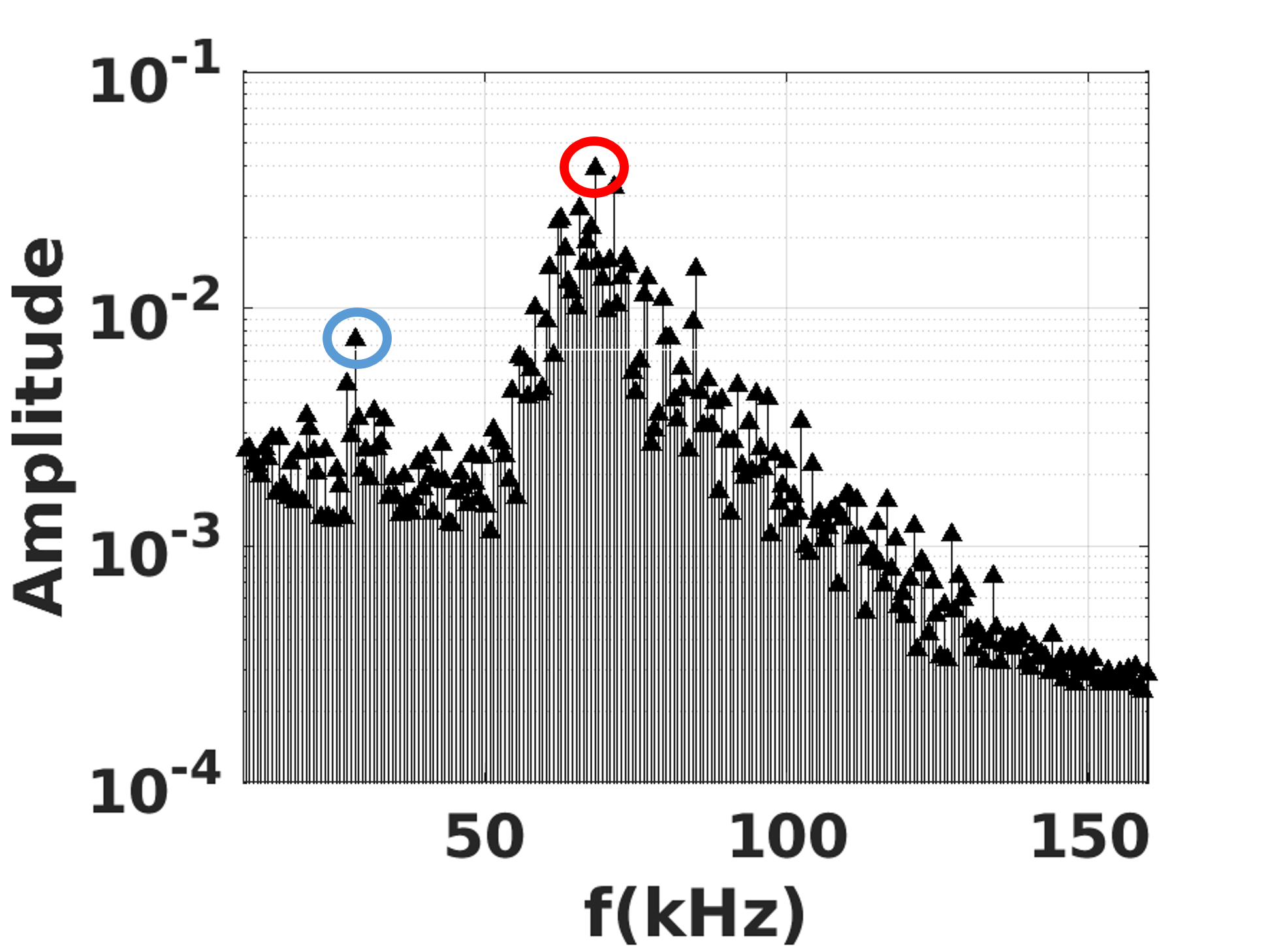}}}
\caption{Frequency-amplitude distribution for linear $3D$ stochastic forcing extracted from DMD analysis (Case~$2$).}
\label{fig:DMD_modes}
\end{figure}

The frequency-amplitude distribution of the pressure perturbation field is shown in Fig.~\ref{fig:DMD_modes}.
The values represent the relative amplification of disturbances in the considered domain and have been extracted with the DMD method which provides a best-fit result in the least-square sense.
$1{,}000$ snapshots separated by $\Delta t= 0.05$ are employed to provide a converged result.
Fluctuations in two frequency ranges,  $f\in[10-50kHz]$ and $f\in[60-100kHz]$, are highlighted.
Based on the numerical neutral curve of Fig.~\ref{fig:Ogv14_stx}, the latter corresponds to second-mode waves, while the former represents first-mode waves as discussed further below.
The most amplified second mode wave~$(f=68.3kHz)$ has an order of magnitude higher amplitude than the corresponding first-mode wave $(f=28.5kHz)$, reinforcing the  dominance of Mack modes in linear disturbance evolution on frustums of ogive-cylinders.

\begin{figure}
\centering
\begin{subfigmatrix}{1}
\subfigure[]{{\includegraphics[trim=5 80 50 100, clip, width=0.75\textwidth]{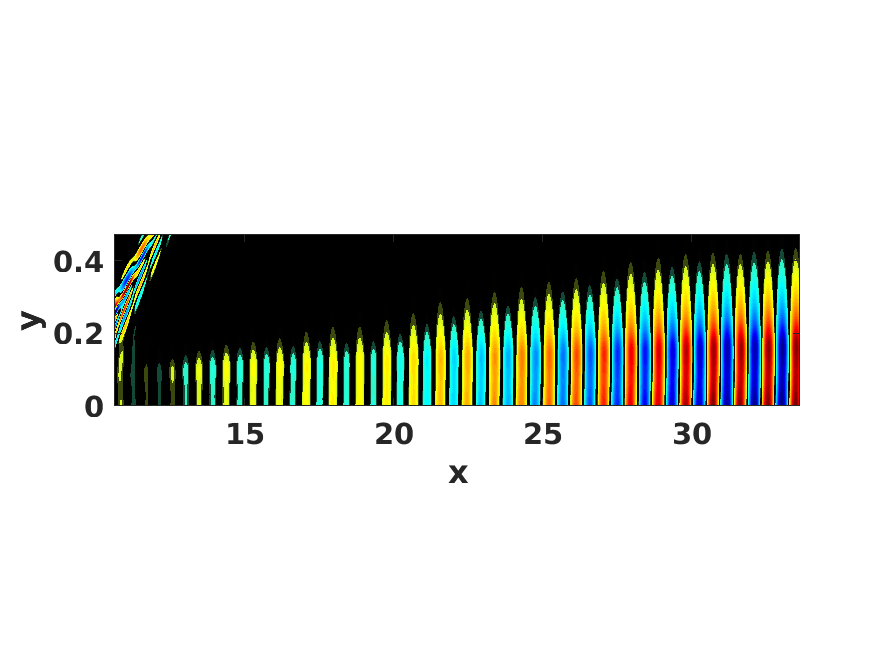}}}
\subfigure[]{{\includegraphics[trim=5 80 50 100, clip, width=0.75\textwidth]{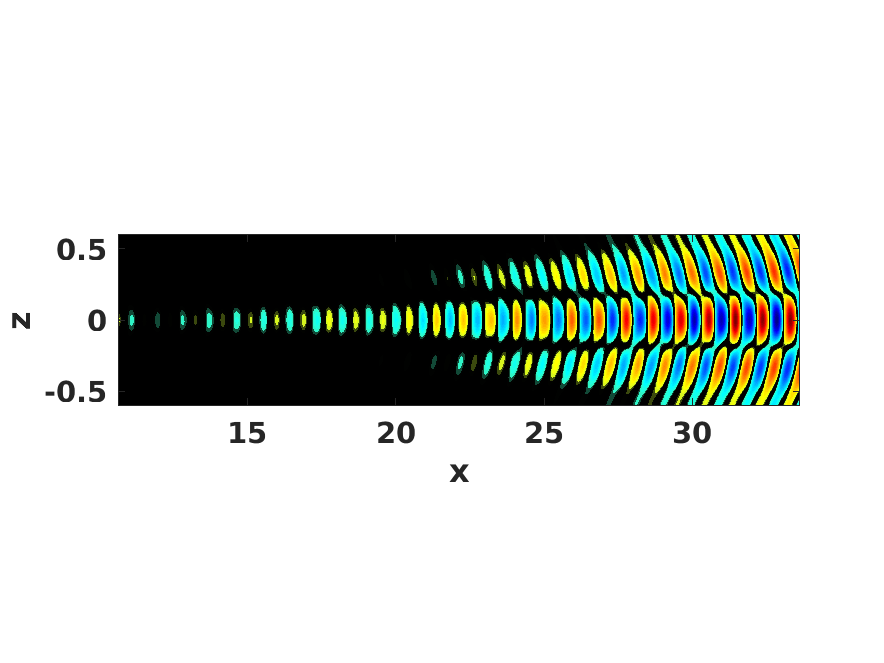}}}
\subfigure[]{{\includegraphics[trim=5 80 40 90, clip, width=0.75\textwidth]{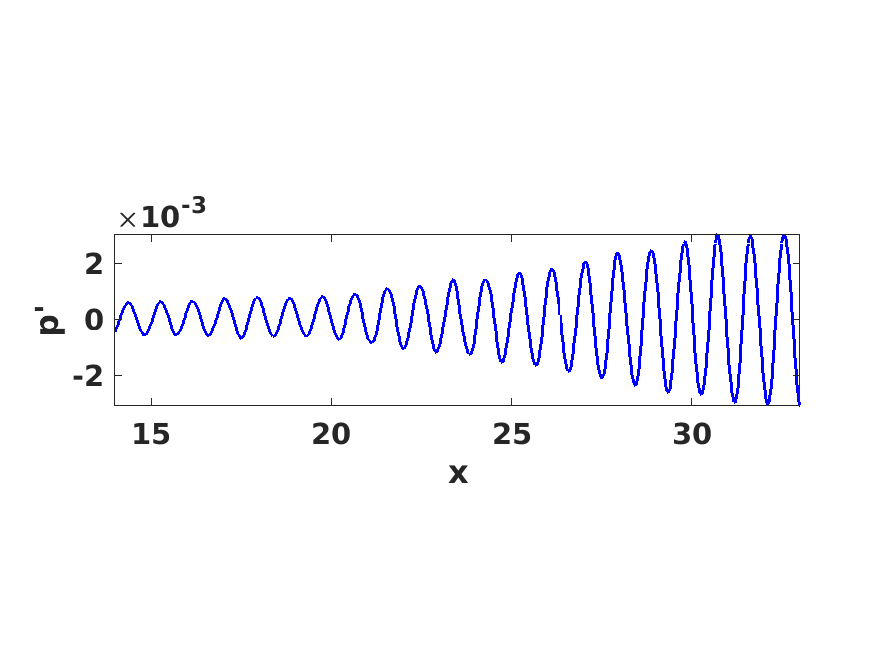}}}
\end{subfigmatrix}
\caption{DMD pressure mode features of first mode at $f=28.55kHz$.
$(a)$ Wall-normal signature at $z=0$.
$(b)$ Spanwise signature at $y=0$.
$(c)$ Amplitude evolution on the line $(y,z)=(0,0)$.}
\label{fig:DMD_FMode}
\end{figure}

The characteristics of the modal structure at $f=28.5kHz$ are shown in Fig.~\ref{fig:DMD_FMode}, with 
contour plots saturated to $80\%$ of their maximum values for ease of interpretation.
The pressure perturbation field (Fig.~\ref{fig:DMD_FMode}(a)) shows single lobe structures in the wall normal direction.
The wall plane data, Fig.~\ref{fig:DMD_FMode}(b), are displayed by unrolling the spanwise cylinder surfaces with the relation $z=r\phi$, where $r$ is the local radius of the cylinder (nondimensional) and $\phi$ is the azimuthal angle.
In this plane, the structures are oblique confirming their first mode nature.
The evolution of the pressure perturbation on the wall at $z=0$ (Fig.~\ref{fig:DMD_FMode}(c)) shows wave amplification in the domain considered.
Other disturbances in the frequency range $f\in[10-50kHz]$ also exhibit similar structures and hence can be classified as first-mode waves (see Mack~\cite{mack1984boundary}).
%\todo{My understanding is that the oblique nature of the first mode was discovered later by someone other than Mack.  Perhaps I'm wrong? HG: Current form is right}

\begin{figure}
\centering
\begin{subfigmatrix}{1}
\subfigure[]{{\includegraphics[trim=5 80 50 100, clip, width=0.75\textwidth]{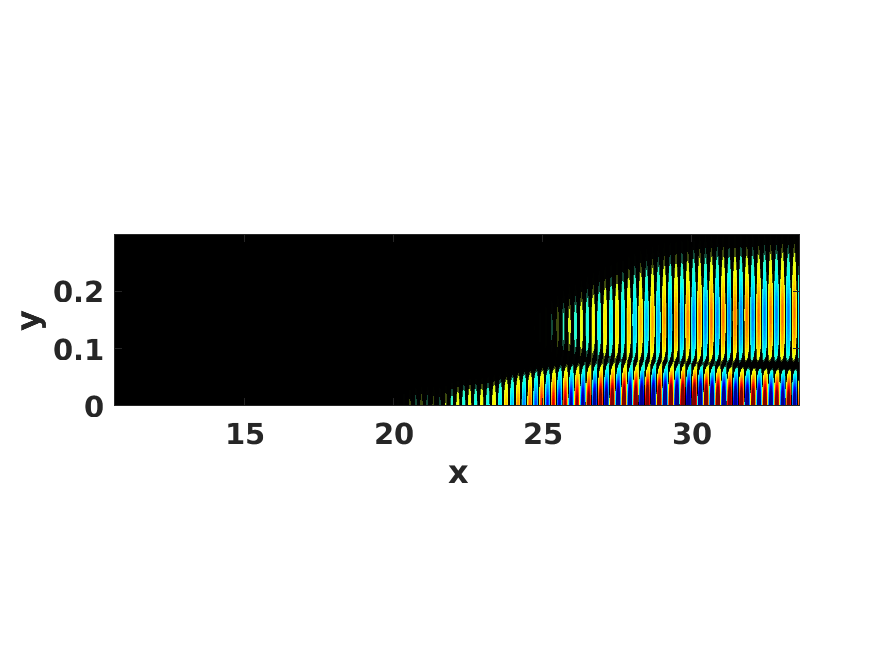}}}
\subfigure[]{{\includegraphics[trim=5 80 50 100, clip, width=0.75\textwidth]{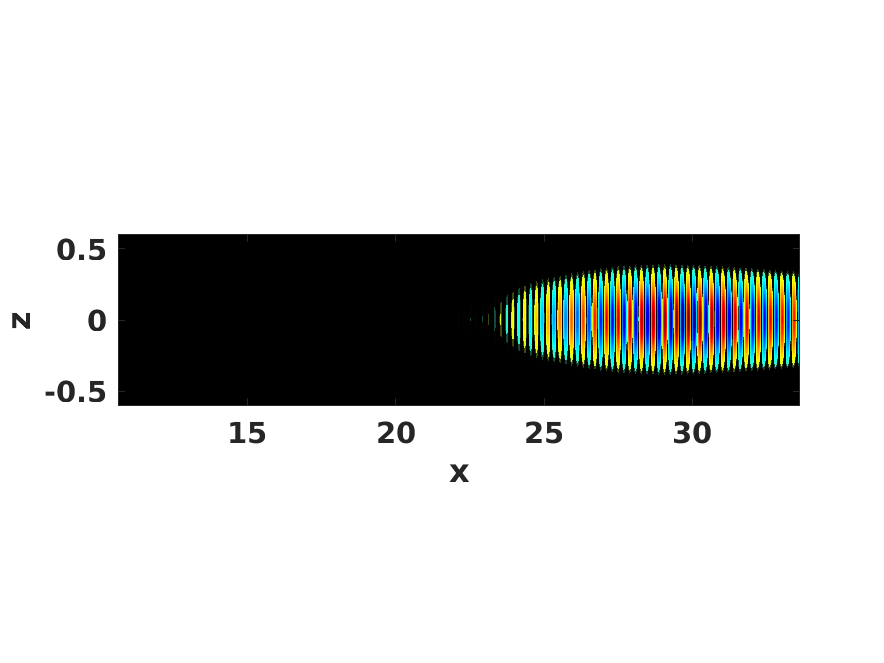}}}
\subfigure[]{{\includegraphics[trim=1 80 40 100, clip, width=0.75\textwidth]{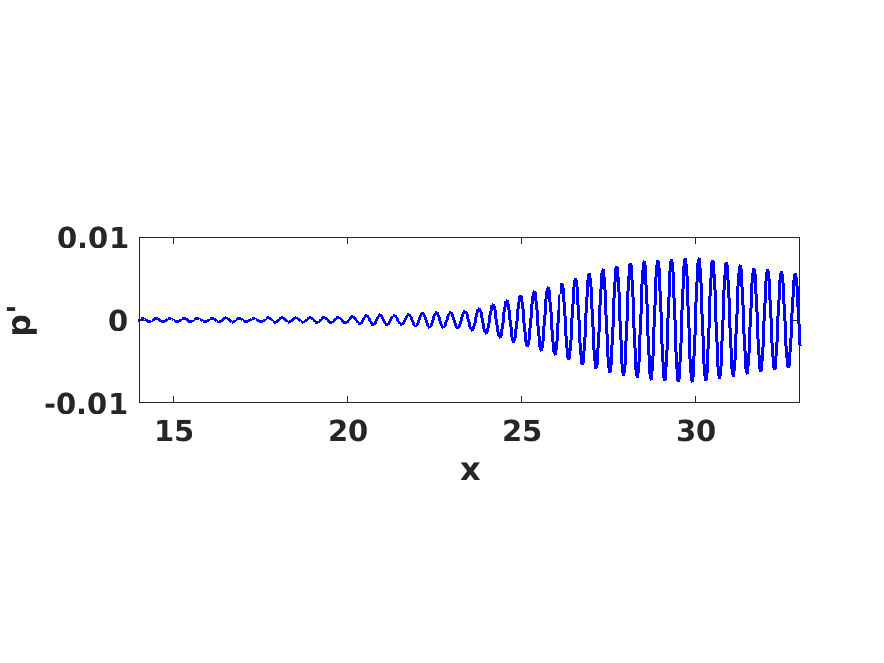}}}
\end{subfigmatrix}
\caption{DMD pressure mode features of second mode at $f=68.3kHz$.
$(a)$ Wall-normal signature at $z=0$.
$(b)$ Spanwise signature at $y=0$.
$(c)$ Amplitude evolution on the line $(y,z)=(0,0)$.}
\label{fig:DMD_SMode}
\end{figure}

The disturbance characteristics of the most amplified wave at $f=68.3kHz$ are depicted in Fig.~\ref{fig:DMD_SMode}. 
Unlike at the lower frequency, in this case, two-lobed structures are evident in the wall normal plane (Fig.~\ref{fig:DMD_SMode}(a)); these are indicative of Mack modes~\cite{mack1984boundary} and their
%\todo{citation would be good. HG:Fixed}.
spanwise signature (Fig.~\ref{fig:DMD_SMode}(b)) is predominantly two-dimensional.
The amplification of these disturbances at the centerline in Fig.~\ref{fig:DMD_SMode}(c) shows relatively larger amplification over a smaller streamwise length than the first-mode wave shown earlier in Fig.~\ref{fig:DMD_FMode}(c).
Other frequencies in $f\in[60-100kHz]$ exhibit similar structures; consistent with the N-factor figure presented earlier, amplitudes of higher frequencies peak at more upstream locations.
A comparison of Figs.~\ref{fig:DMD_FMode} and~\ref{fig:DMD_SMode} indicates that first-mode waves have longer streamwise wavelengths than the second modes.
While the former amplify over a longer streamwise extent, the latter exhibit larger amplification over a relatively smaller domain.
%These inferences suggest that the lower frequency disturbances do not reach $FS$~synchronization in the current domain. 
%???the above sentence is not clear.
%HG: Removed the sentence.%Higher frequency waves can be considered as $S$-modes post synchronization (see,~\cite{wang2009effect}).

\subsection{Propagation of Three-dimensional Wave Packets}
\label{sec:WP}
\subsubsection{Linear Disturbance Evolution}
\begin{figure}
\begin{subfigmatrix}{3}
\subfigure[]{{\includegraphics[width=0.33\textwidth]{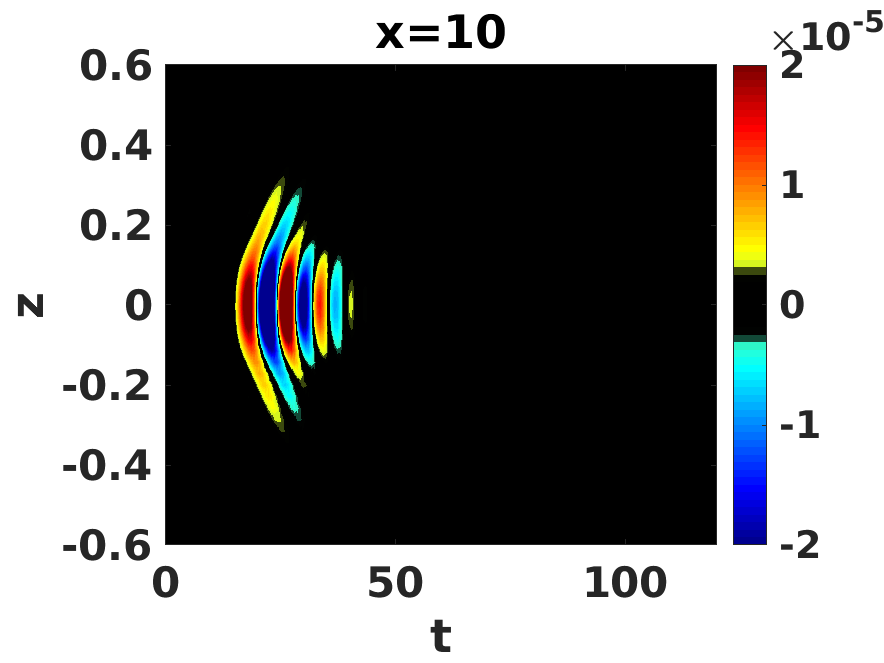}}}
\subfigure[]{{\includegraphics[width=0.33\textwidth]{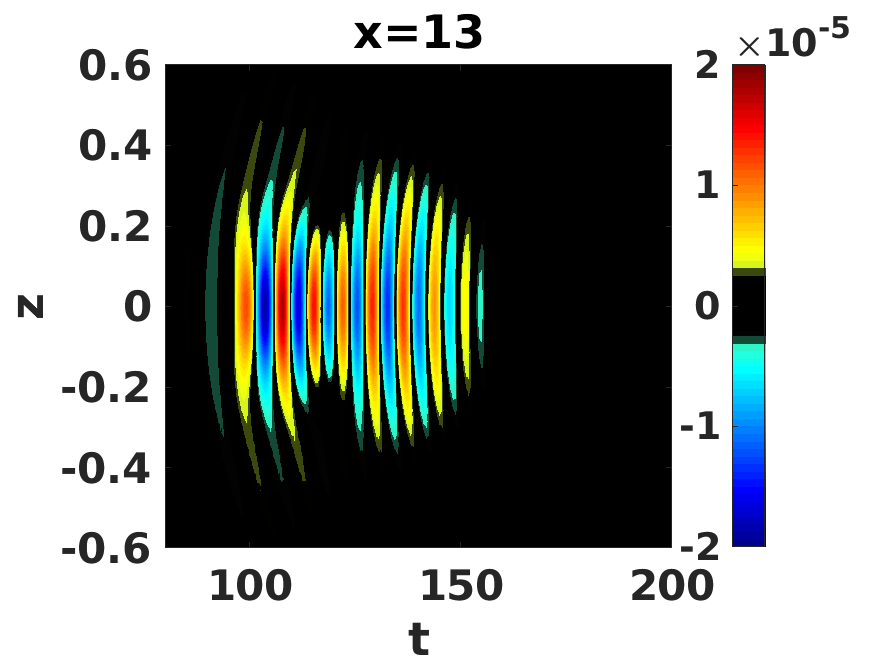}}}
\subfigure[]{{\includegraphics[width=0.33\textwidth]{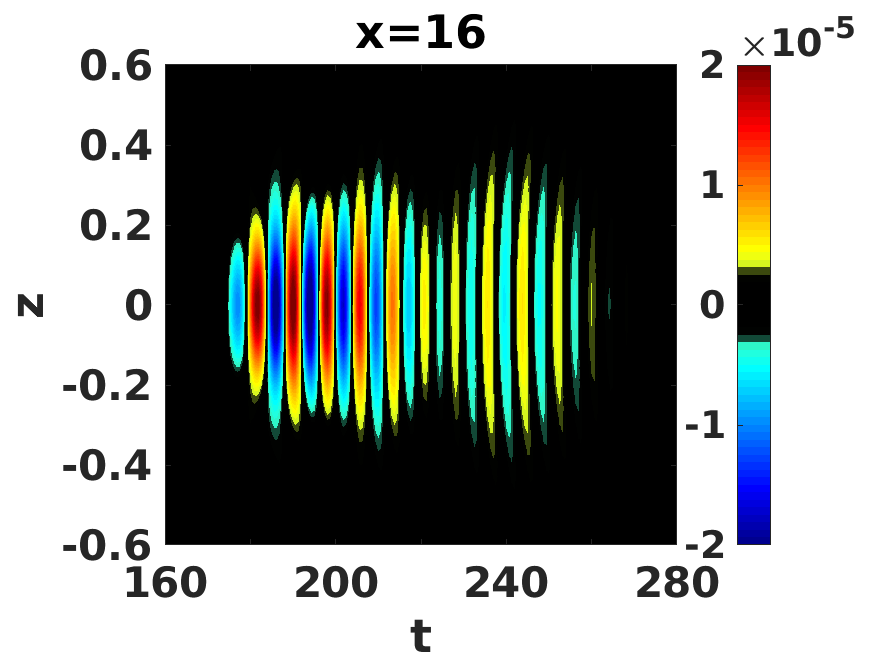}}}
\subfigure[]{{\includegraphics[width=0.33\textwidth]{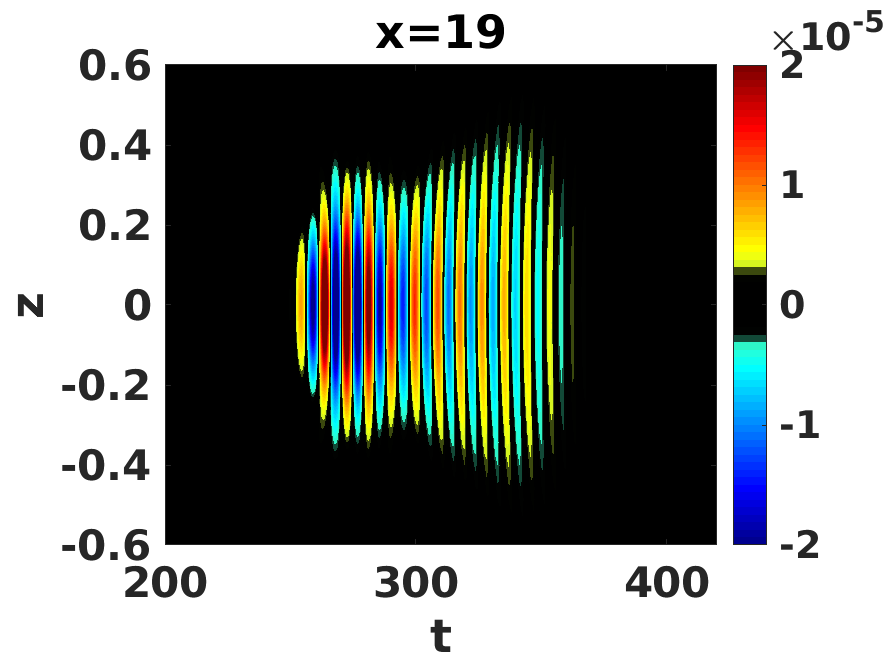}}}
\subfigure[]{{\includegraphics[width=0.33\textwidth]{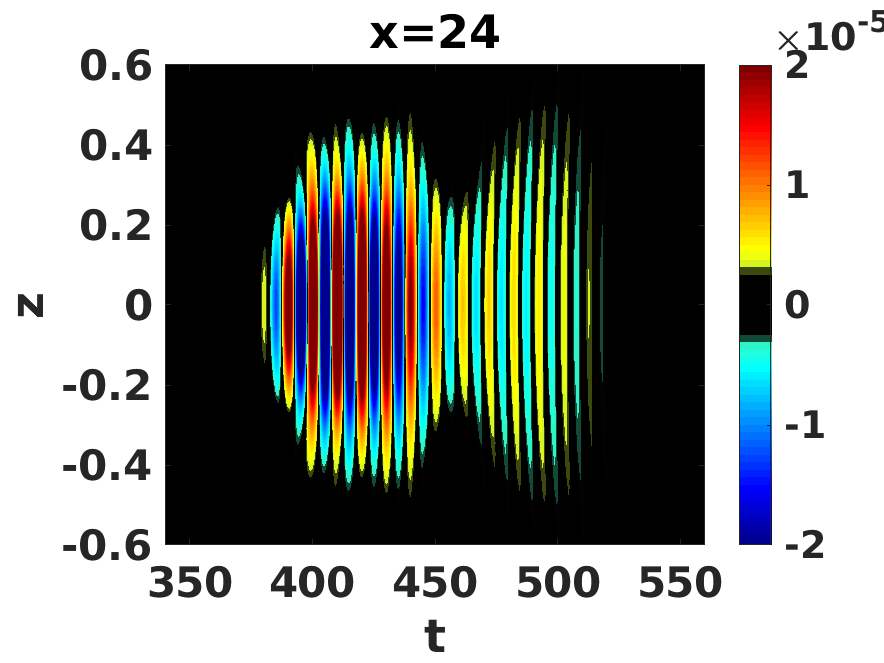}}}
\subfigure[]{{\includegraphics[width=0.33\textwidth]{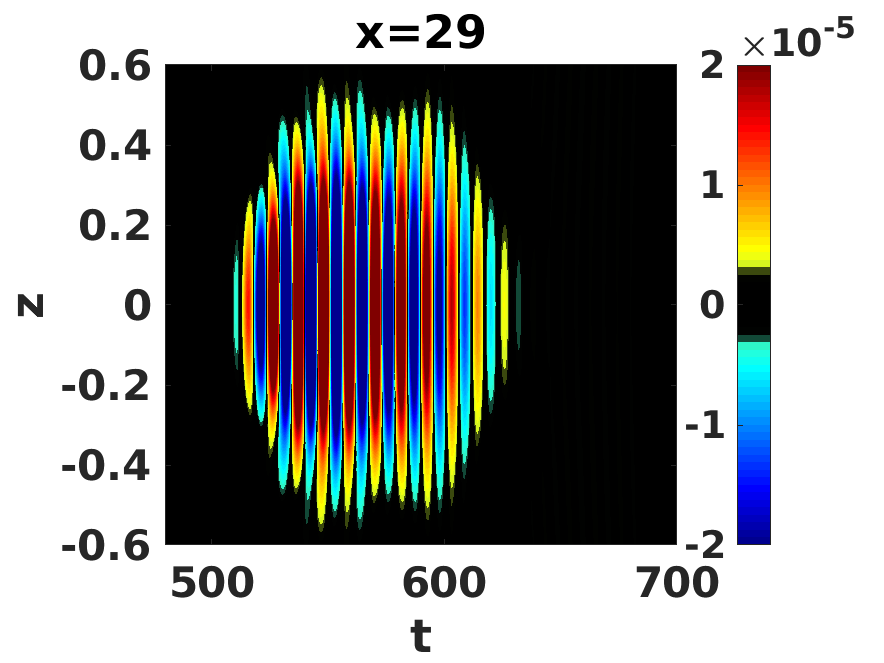}}}
\end{subfigmatrix}
\caption{Spatiotemporal evolution of wall pressure perturbations for linear wave packet forcing (Case~$3$).}
\label{fig:LFM_PP}
\end{figure}

To examine the three-dimensional features of linear disturbance amplification, a small-amplitude wave packet~$(\epsilon=5\times 10^{-4})$  centered at $x=9.46$ (Case~$3$) is triggered on the wall.
This location is relatively closer to the ogive-cylinder junction, to exploit the fact that the natural receptivity is higher in this region.
Figure~\ref{fig:LFM_PP} illustrates the spatiotemporal evolution of the wave packet disturbance (instantaneous minus steady laminar basic state) through the wall pressure perturbation signature at various streamwise locations.
%The disturbance fields are obtained by subtracting the steady laminar base flow from the instantaneous flow fields. 
Since the horizontal axis represents non-dimensional time, the leftmost point on the contour corresponds to the signature of the leading edge of the wavepacket at the indicated streamwise location.

Initially, disturbances emanating from the forcing area ($x=10$, Fig.~\ref{fig:LFM_PP}(a)) have leading circular arc-like wave fronts.  
%\todo{``semicircular would be -90 to 90''. Can we say circular arc-like? HG:Done}
As the disturbances evolve ($x=13$, Fig.~\ref{fig:LFM_PP}(b)) however, arrowhead shaped structures develop behind these wave fronts.
Further downstream ($x=16$, Fig.~\ref{fig:LFM_PP}(c)), the wave packet elongates in the streamwise axis and modulates in the spanwise extent.
This structure resembles a flat plate wave packet, with oblique disturbances spread at the trailing edge~\cite{sivasubramanian2011transition}; on cones, the body curvature results in the formation of oblique waves at the center of the wave packet~\cite{sivasubramanian2015direct}.
In the present case, the wave packet also develops through spreading at the trailing edge in the streamwise and azimuthal extents, as seen in $x=19$ in Fig.~\ref{fig:LFM_PP}(d).
%???the above sentence is ambiguous - what do you mean by spreading at trailing edge? is it streamwise elongation or azimuthal extent?
%HG: In both directions- now clarified
The relative amplitudes of the disturbances are higher in the leading arrowhead portion relative to the trailing waves, which is especially evident in 
Figs.~\ref{fig:LFM_PP}(e,f)~$(x=24,~26)$.
%show that these leading disturbances amplify more than the trailing waves.
Furthermore, they spread more in the streamwise rather than in the azimuthal direction.
%\todo{How do we discern this from the figures? HG:I replaced streamwise direction to axis- to note the x-axis (time) is similar to streamwise direction} 

\begin{figure}
\centering
\begin{subfigmatrix}{3}
\subfigure[]{{\includegraphics[width=0.33\textwidth]{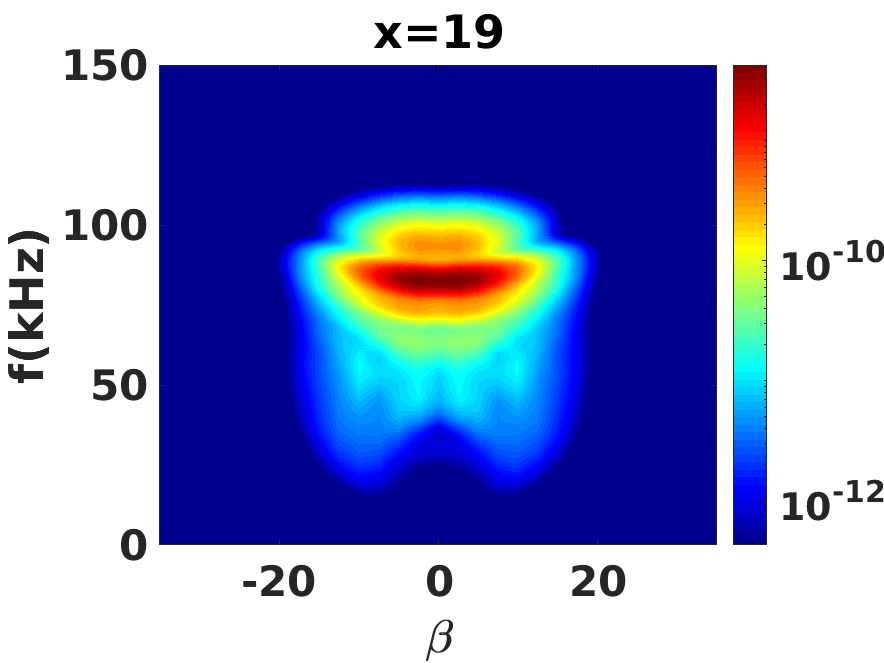}}}
\subfigure[]{{\includegraphics[width=0.33\textwidth]{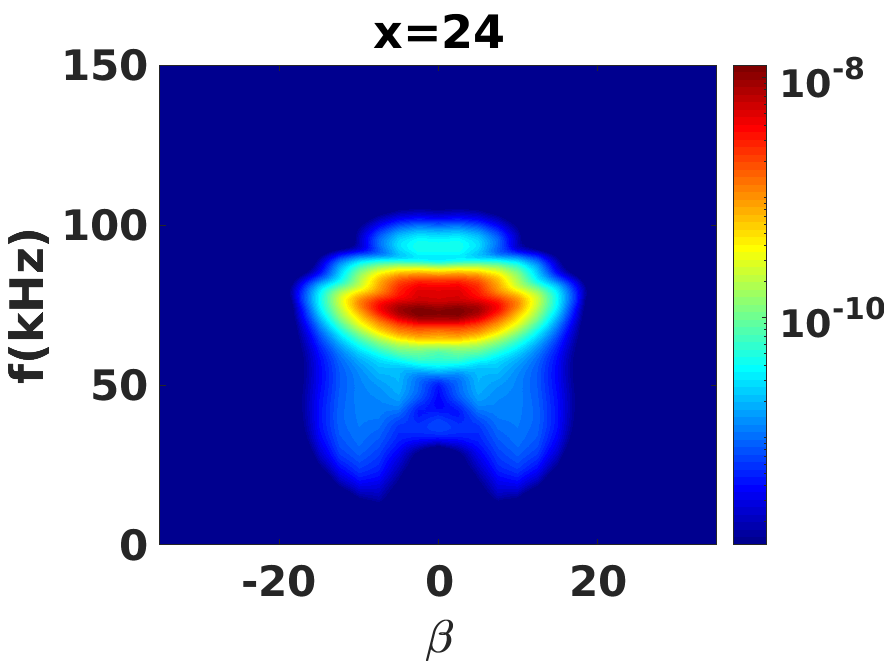}}}
\subfigure[]{{\includegraphics[width=0.33\textwidth]{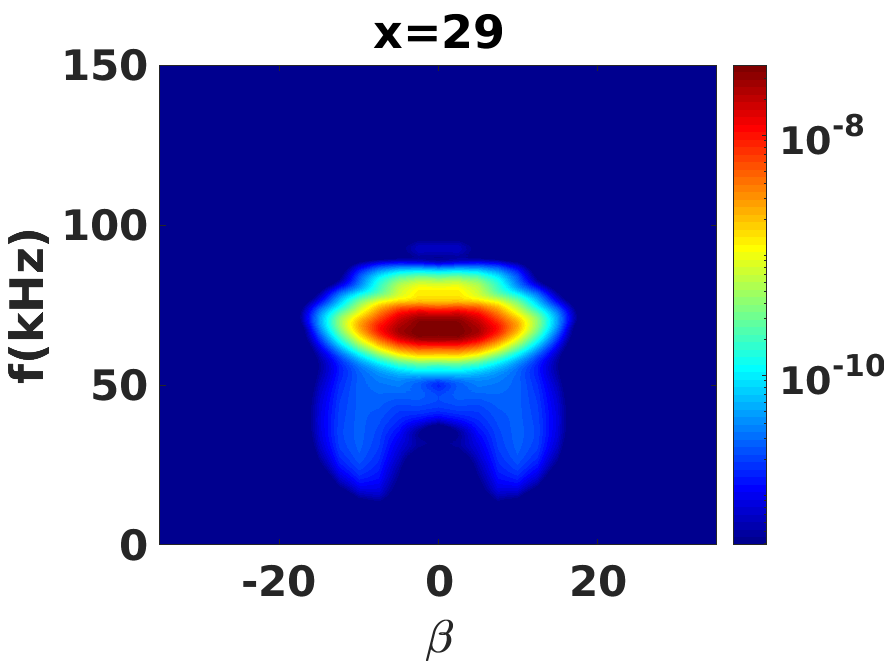}}}
\end{subfigmatrix}
\caption{Streamwise spectral evolution contours of the linear wave packet in the frequency $(f)$-spanwise wavenumber $(\beta)$ plane (Case~$3$).}
\label{fig:LFM_FFT}
\end{figure}

Two-dimensional spanwise wavenumber-frequency~$(\beta-f)$ spectra of these disturbances, extracted by applying $2D$ Fourier transformation on the unrolled spanwise coordinates, are shown in Fig.~\ref{fig:LFM_FFT}.
The spectral signature at $x=19$ in Fig.~\ref{fig:LFM_FFT}(a) shows that two-dimensional~$(\beta=0)$ high-frequency disturbances~$(f\sim82kHz)$ have larger amplitudes than their oblique low-frequency~$(f\sim 22kHz)$ counterparts.
The higher and lower amplitude waves correspond to the leading $2D$ Mack modes and the trailing first modes, respectively.
The peak $2D$ frequency decreases downstream (Fig.~\ref{fig:LFM_FFT}(b,c)) ($f\sim 70.5kHz$,~$65.2kHz$, respectively), consistent with the inverse scaling of Mack modes with the thickness of the boundary layer~\cite{marineau2019analysis}.
The signature of the leading-wave amplification in Fig.~\ref{fig:LFM_PP}(e,f) is also apparent in Fig.~\ref{fig:LFM_FFT}(b,c).
Therefore, Mack modes dominate linear instabilities in this configuration, consistent with the stability analysis-based observations of Scholten et al.~\cite{scholten2022linear}.
%\todo{Need to make sure that the distinction between what Scholten did and what this paper is doing comes through clearly.  If they did stability, while we're doing DNS, thsi should be implicitly brought out by just putting ``stability analysis-based observations of Scholten ...'' or something like that. HG: Added "stability-based analysis"}
 
\begin{figure}
\centering
\begin{subfigmatrix}{3}
\subfigure[]{{\includegraphics[width=0.33\textwidth]{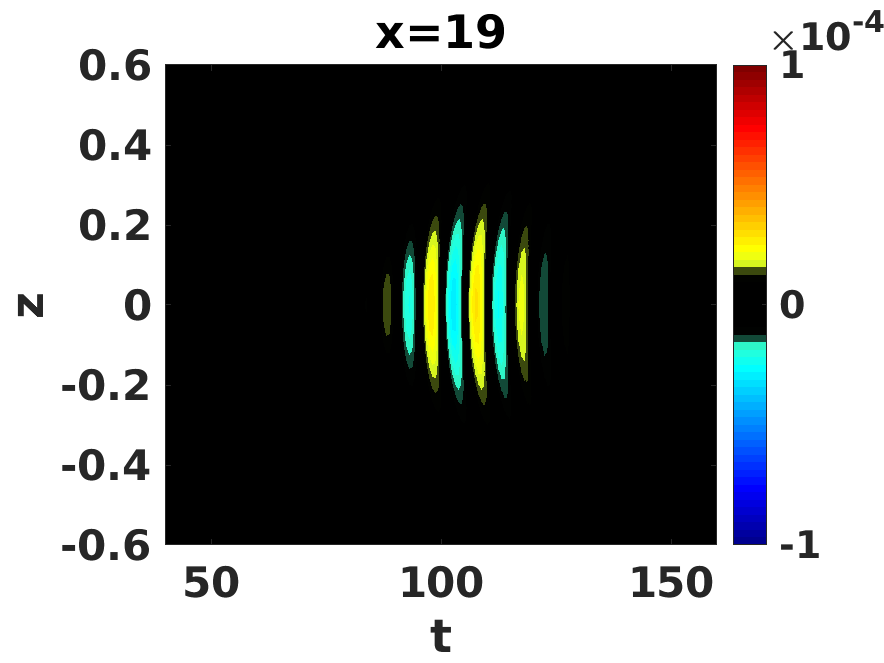}}}
\subfigure[]{{\includegraphics[width=0.33\textwidth]{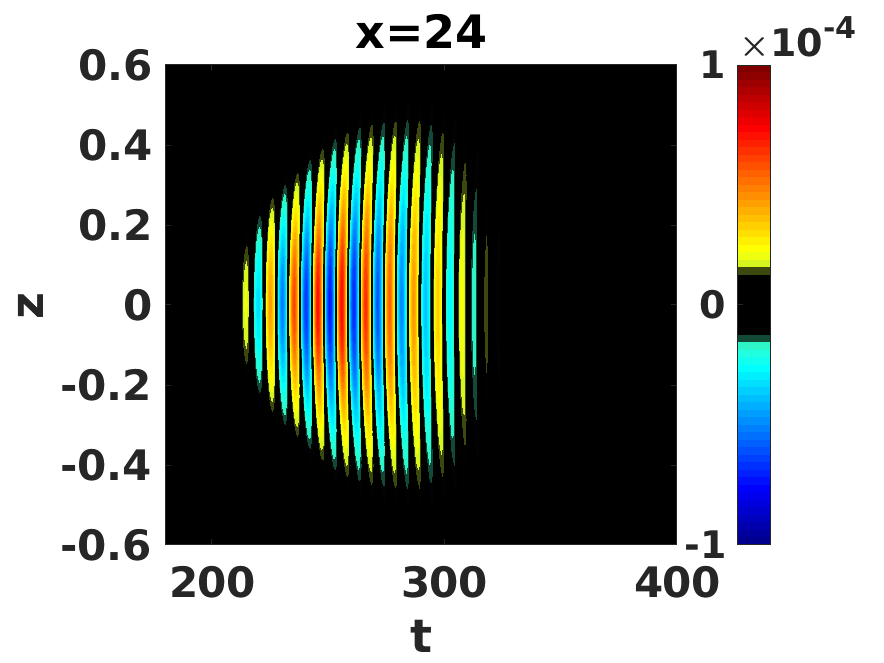}}}
\subfigure[]{{\includegraphics[width=0.33\textwidth]{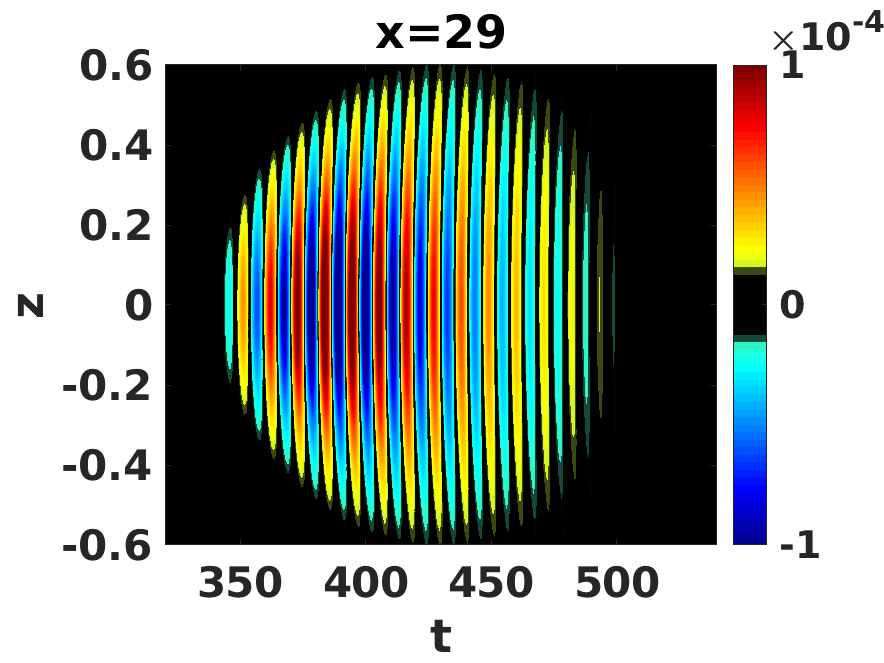}}}
\subfigure[]{{\includegraphics[width=0.33\textwidth]{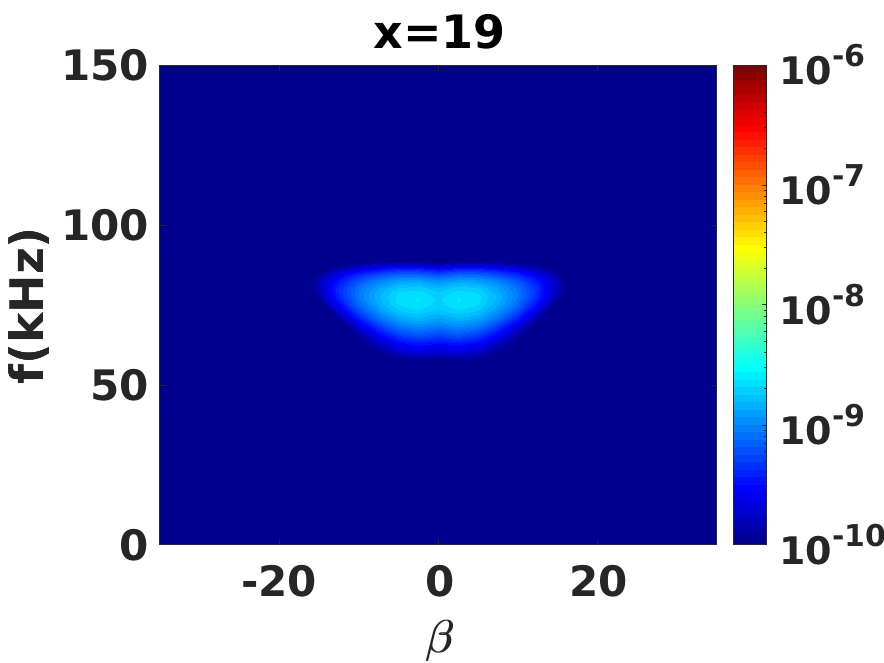}}}
\subfigure[]{{\includegraphics[width=0.33\textwidth]{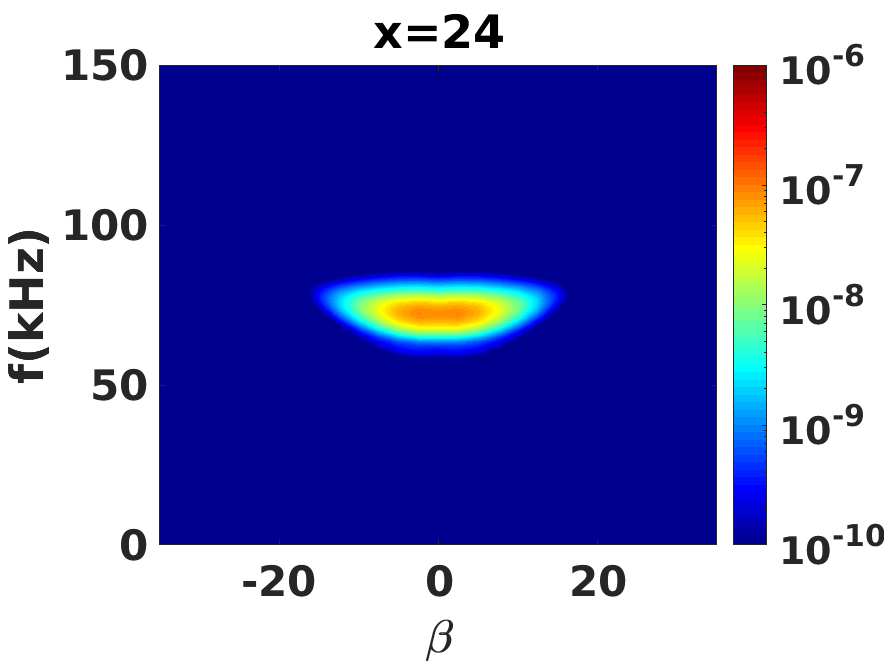}}}
\subfigure[]{{\includegraphics[width=0.33\textwidth]{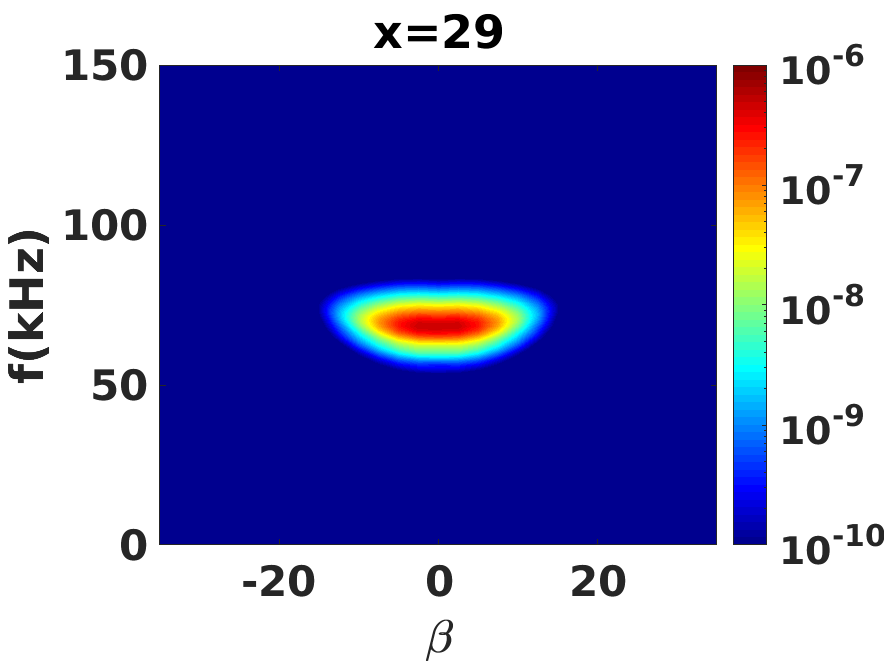}}}
\end{subfigmatrix}
\caption{Linear wave packet evolution for disturbances triggered at $x=16.2$ (Case~$4$).
Wall pressure perturbation contours are shown in $(a,b,c)$ while the
The corresponding spectral content is plotted in $(d,e,f)$.}
\label{fig:LSM_PP}
\end{figure}

The effect of actuator location on instability evolution is examined by considering excitation at a downstream location of $x=16.2$ (Case~$4$).
The wall pressure perturbation signature along with its $\beta-f$ spectral distribution are shown in Fig.~\ref{fig:LSM_PP}.
At $x=19$, Fig.~\ref{fig:LSM_PP}(a), the wave packet displays an arrowhead shape with a predominantly $2D$ signature based on the wavenumber content of Fig.~\ref{fig:LSM_PP}(d).
These arrowhead structures develop further downstream (Fig.~\ref{fig:LSM_PP}(b,c)) with no apparent signature of the trailing waves observed at the upstream actuator location (Fig.~\ref{fig:LFM_PP}).
The corresponding spectral distribution (Fig.~\ref{fig:LSM_PP}(e,f))) captures Mack modes with decreasing frequency ($f\sim 70.5kHz$,~$65.2kHz$, respectively) and increasing amplitudes, similar to those in Fig.~\ref{fig:LFM_FFT}.
However, the oblique wave signature is not observed in the Fourier plane.

The above wave packet analyses suggest that disturbance evolution depends on the location of actuation.
In free flight, one possible source of such actuation may be particle impacts at different  locations on the surface of the vehicle, with  amplitude disturbance depending on the size of the particle~\cite{browne2021numerical}.
Similar changes in receptivity have previously been observed using wave packets on sharp cones~\cite{hader2021three}, compression corners~\cite{novikov2016direct}, and blunted flat plates~\cite{goparaju2021role,scholten2022plate}.
On hypersonic sharp flat plates, Sivasubramanian et al.~\cite{sivasubramanian2016reynolds} attributed the differences in receptivity to the difference in flow sensitivity to the first and second discrete modes at different actuator locations.

\begin{figure}
\centering
{\includegraphics[trim=5 60 10 25,clip,width=0.75\textwidth]{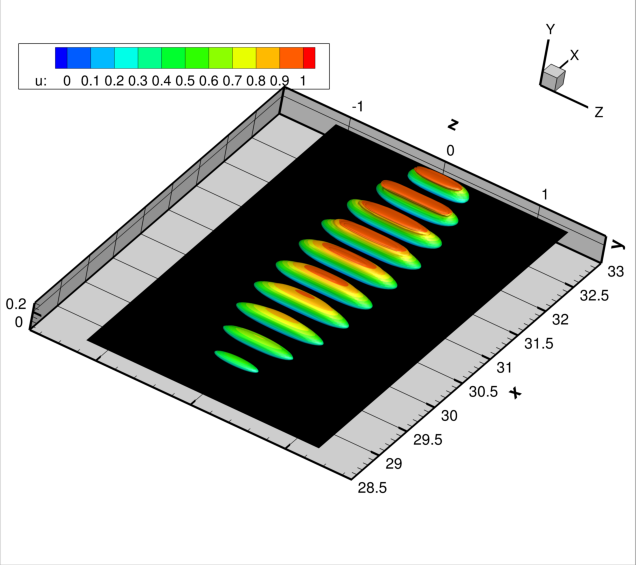}}
\caption{Instantaneous vortical structures identified by the Q-criterion $(Q=0.01)$ isosurfaces extracted at the end of the domain for Case~$3$.}
\label{fig:L_Qcrit}
\end{figure}

\begin{figure}
\centering
{\includegraphics[trim=0 40 10 60, clip, width=0.75\textwidth]{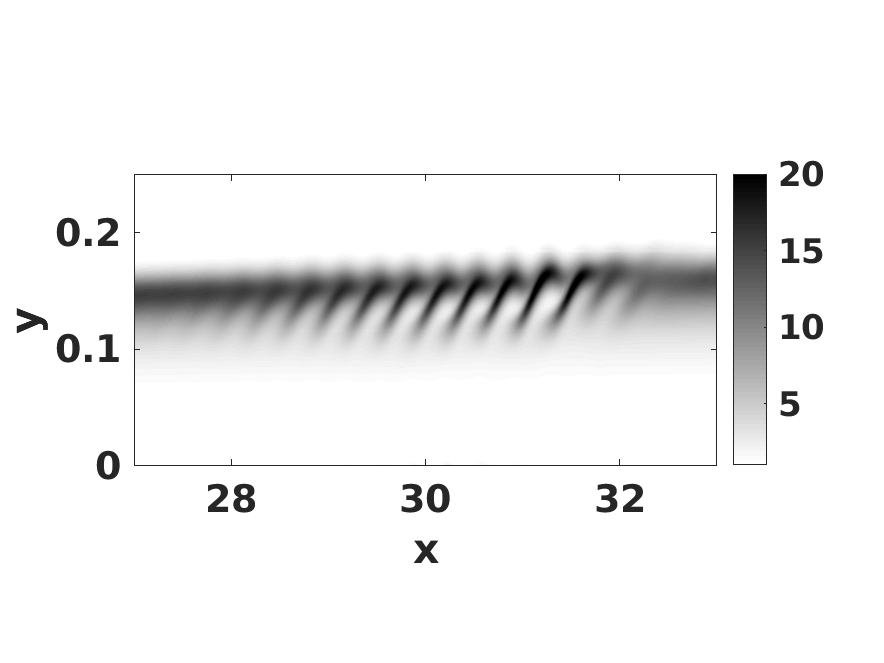}}
\caption{Pseudo-Schlieren at the symmetry plane of the instantaneous vortical structures in Case~$3$.
}
\label{fig:L2D_gradrho}
\end{figure}

\begin{figure}
\centering
{\includegraphics[trim=0 0 0 0,clip,width=0.75\textwidth]{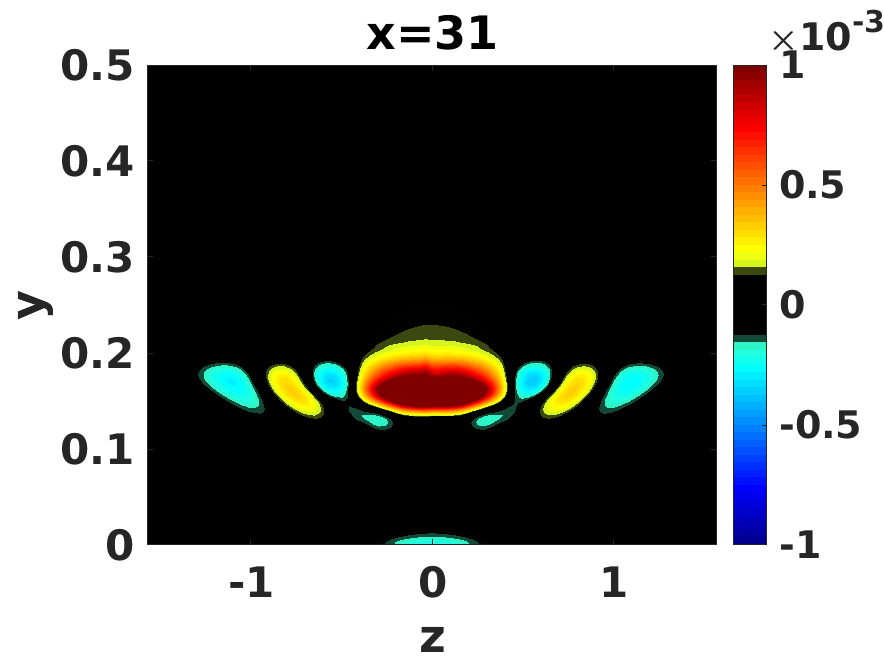}}
\caption{Density perturbation contours in the cross-flow plane of instantaneous vortical structures (Case~$3$).}
\label{fig:LinQR}
\end{figure}

The vortical structures manifested by Case~$3$ wave packet excitation are examined with $Q$-criterion~\cite{hunt1988eddies} isosurfaces closer to the exit of the domain in Fig.~\ref{fig:L_Qcrit}.
Two-dimensional roller-like structures typical of Mack modes are observed; consistent with wall pressure perturbation patterns, the oblique first-mode signature is not apparent. 
The wall-normal density gradient structures (pseudo-Schlieren) of this wave packet (at $z=0$) shown in Fig.~\ref{fig:L2D_gradrho} exhibits \textit{rope-like} structures.
The maximum density gradient appears at the boundary-layer GIP (generalized inflection point~\cite{lees1946investigation}) at~$y\sim0.124$, with a nondimensional streamwise wavelength~$0.5$, resulting in a phase speed of~$0.836$.
%???can you confirm if the above corresponds to the boundary layer edge or the GIP?
%%HG: It is more pronounced at GIP- now edited
This is typical of Mack modes and was also reported by the experiments of Hill et al.~\cite{luke2022experimental}.
The density perturbation field in the cross-flow plane at $x=31$ (Fig.~\ref{fig:LinQR}) shows structures near the wall $(y,z)\sim(0,0)$, and at the edge of the boundary layer $(y,z)\sim(0.2,0)$.  
The rope-like structures in the (experimental) Schlieren are manifestations of these stronger density perturbations, occurring away from the wall.
Oblique waves with weaker amplitudes are also observed here with $\lambda_z\sim 0.62$ and correspond to those in Fig.~\ref{fig:LFM_FFT}(c).
%In the next Section, nonlinear evolution features of the wave packets are discussed.

\subsubsection{Nonlinear Modulation of Wave Packet}
\begin{figure}
\centering
\begin{subfigmatrix}{3}
\subfigure[]{{\includegraphics[width=0.33\textwidth]{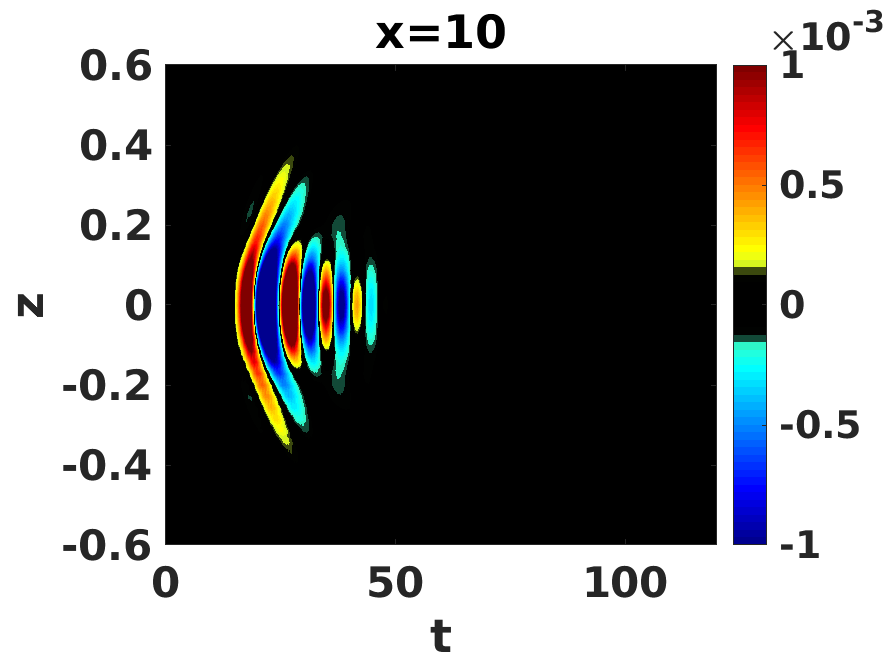}}}
\subfigure[]{{\includegraphics[width=0.33\textwidth]{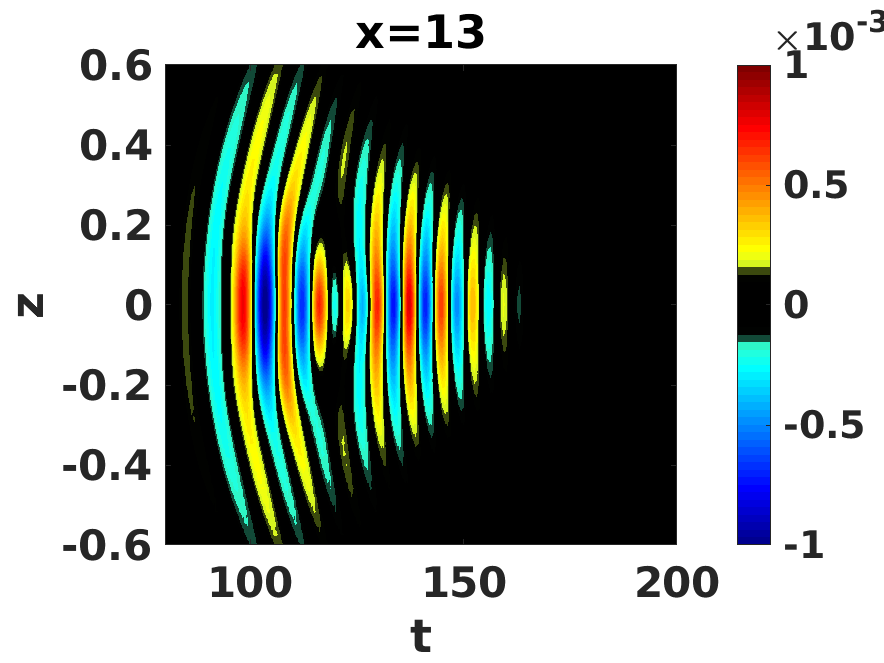}}}
\subfigure[]{{\includegraphics[width=0.33\textwidth]{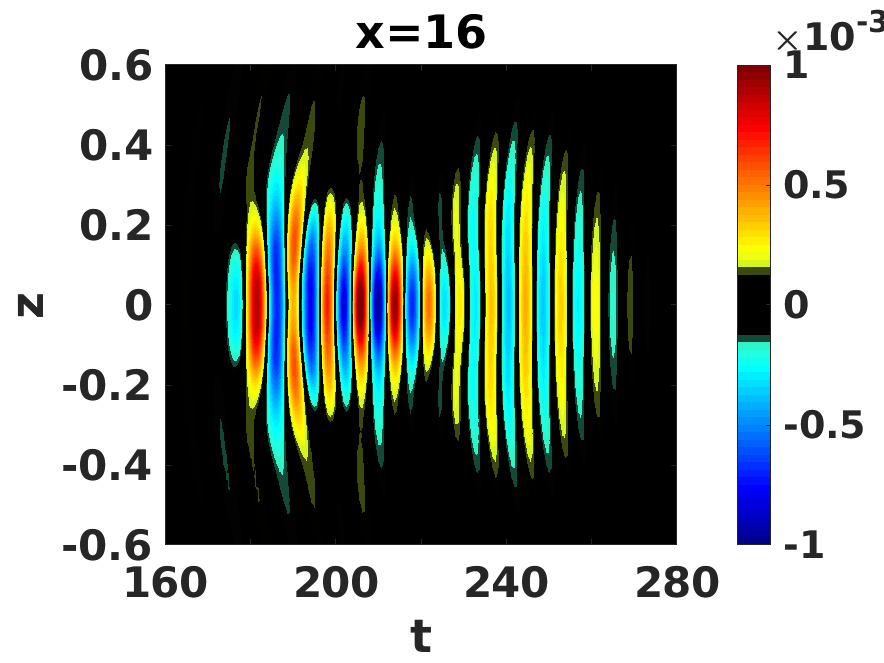}}}
\subfigure[]{{\includegraphics[width=0.33\textwidth]{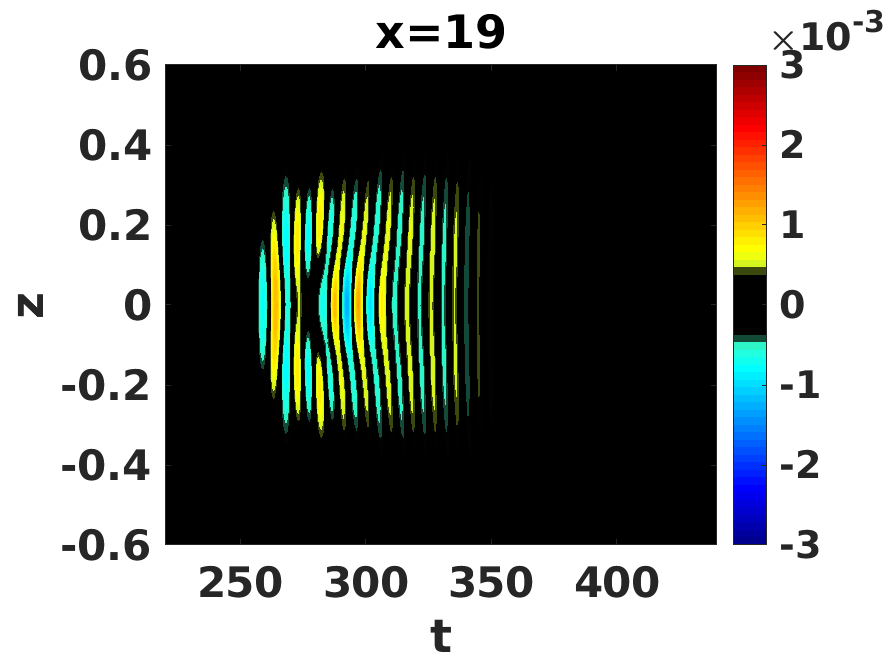}}}
\subfigure[]{{\includegraphics[width=0.33\textwidth]{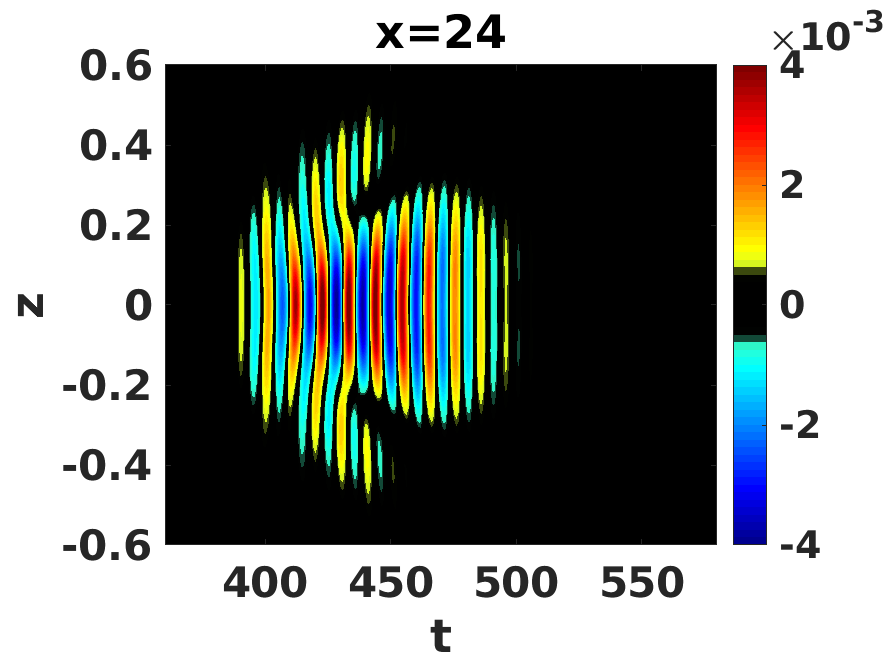}}}
\subfigure[]{{\includegraphics[width=0.33\textwidth]{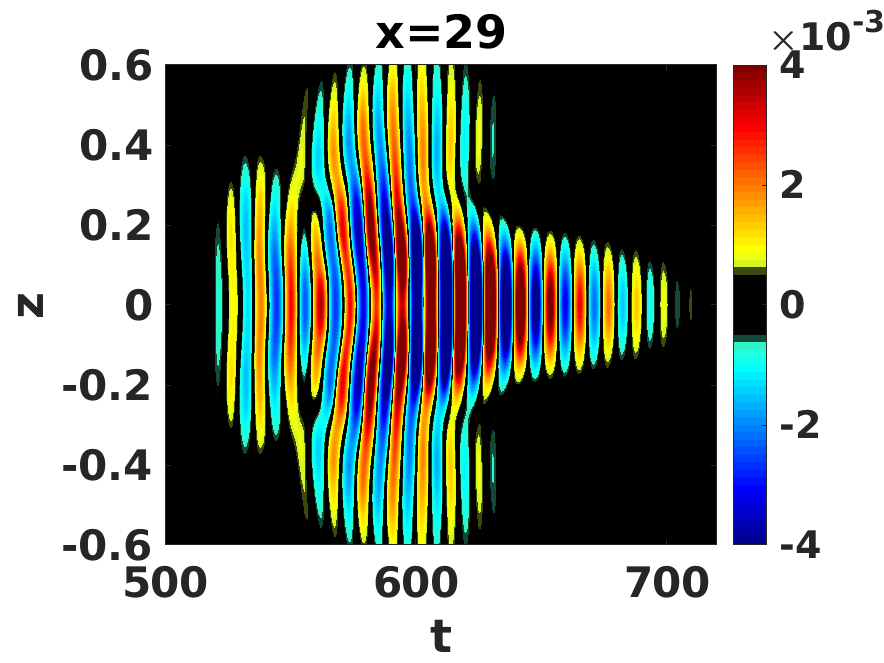}}}
\end{subfigmatrix}
\caption{Spatiotemporal evolution of wall pressure perturbations for nonlinear wave packet forcing (Case~$5$).}
\label{fig:NLFM_PP}
\end{figure}

To examine the effects of nonlinearity, the forcing amplitude is increased to $\epsilon=5\times10^{-2}$ and a wave packet is triggered close to the ogive-cylinder junction at $x=9.46$ (Case~$5$).
The wall pressure perturbation signature is shown in Fig.~\ref{fig:NLFM_PP}. 
Note that the amplitude scales have been adjusted in the subfigures to highlight the dominant features. 
Although the linear wave packets on the ogive-cylinders exhibit structures similar to those on the flat plates, the nonlinear behavior is clearly different.
The signature of the initial disturbance development (Figs.~\ref{fig:NLFM_PP}(a-c)) is similar to the weak forcing case in Fig.~\ref{fig:LFM_PP}(a-c).
Circular arc-like waves constitute the leading front of the wave packet, followed by arrowhead-shaped structures in Fig.~\ref{fig:NLFM_PP}(a,b).
At $x=16$ (Fig.~\ref{fig:NLFM_PP}(c)), the dominance of $2D$ waves at the center of wave packet surrounded by azimuthally varying structures at the trailing end is evident.

Deviations from the weak forcing case become more pronounced further downstream.
%wave packet development. 
The azimuthal modulation of the wave fronts is noticeable in Fig.~\ref{fig:NLFM_PP}(d) and becomes more pronounced further downstream (Fig.~\ref{fig:NLFM_PP}(e)).
The leading portion spreads over the span; however, the arrowhead structure with dominance of $2D$ waves is preserved. 
%\todo{Seems like there's a dip of sorts at x=19 -- the ranges are different for the different subfigures.  Perhaps this should be mentioned so there's no question of comparing amplitudes. HG:Fixed}
Disturbances spread further in the azimuthal direction, maintaining the three-legged structure of the wave packet (Fig.~\ref{fig:NLFM_PP}(f)). 
These structures could be attributed to the significant spanwise curvature in ogive-cylinder bodies, relative to the relevant instability wavelengths.
In sharp cones, such structures were observed when forcing is imposed closer to the tip of the nose, but not for an actuation further downstream on the frustum~\cite{hader2021three}. 
This is not unanticipated because cones have a diverging shape downstream, and the spanwise curvature effects are more prominent closer to the nose than the base.

\begin{figure}
\centering
\begin{subfigmatrix}{3}
\subfigure[]{{\includegraphics[width=0.33\textwidth]{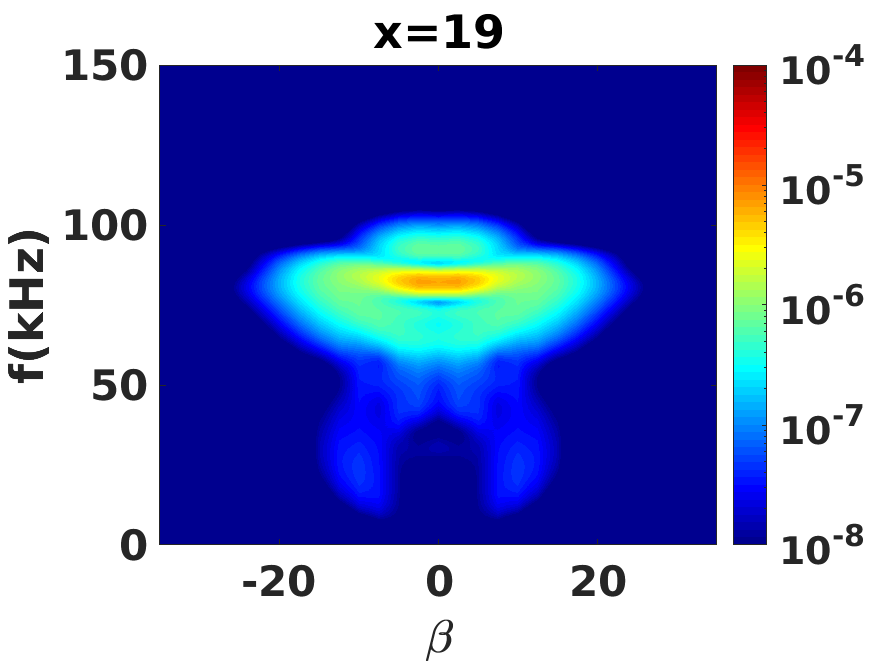}}}
\subfigure[]{{\includegraphics[width=0.33\textwidth]{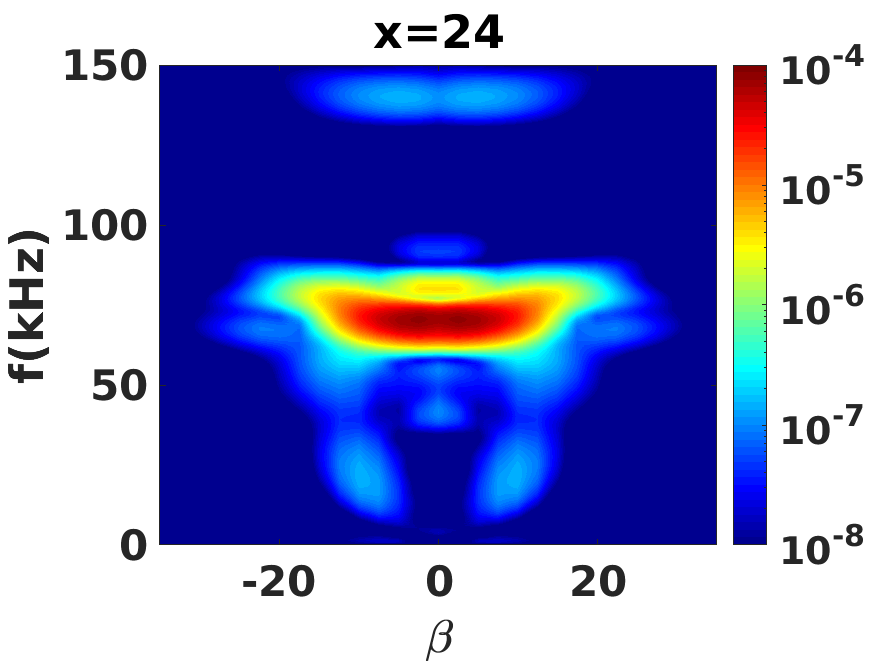}}}
\subfigure[]{{\includegraphics[width=0.33\textwidth]{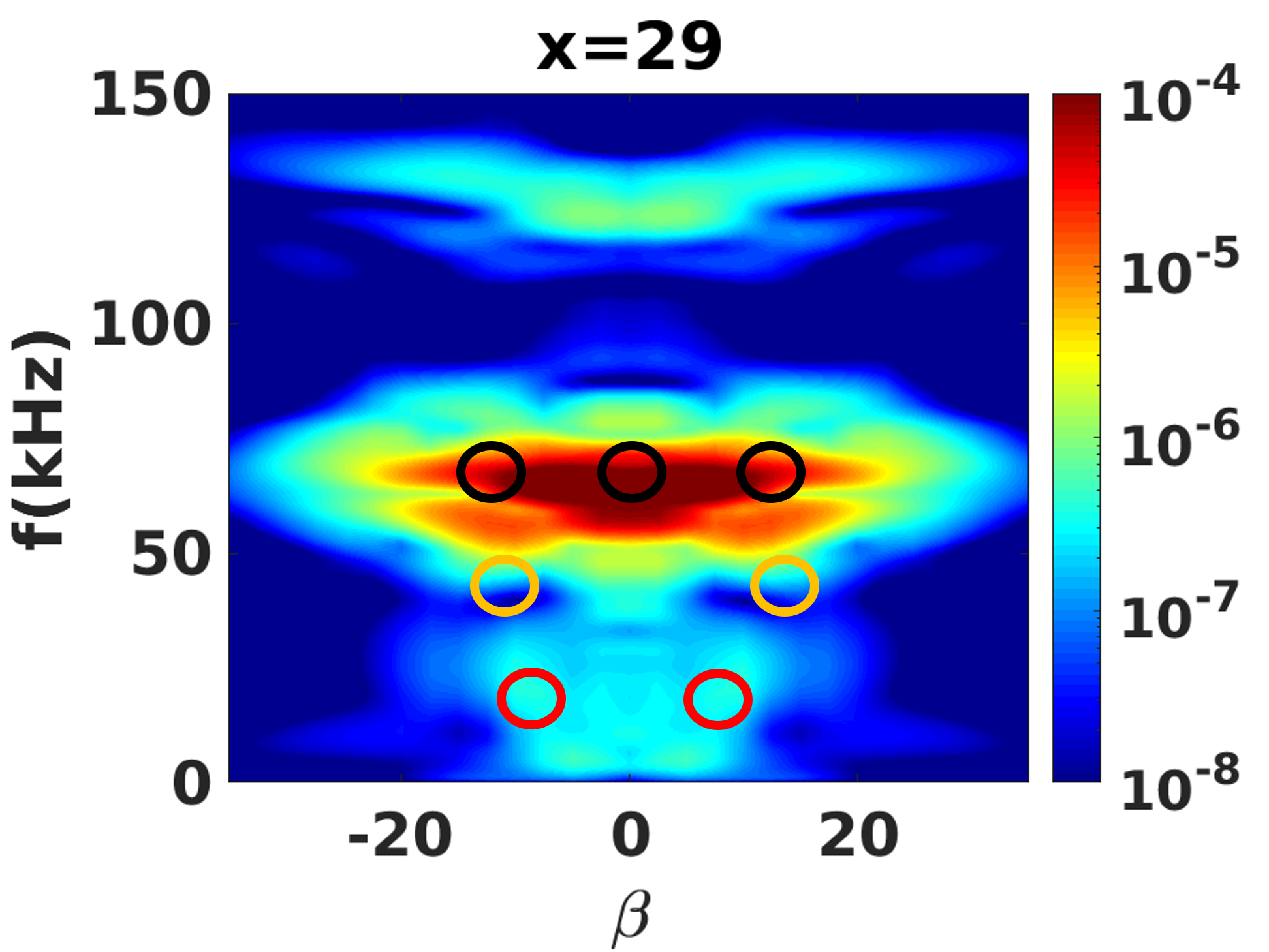}}}
\end{subfigmatrix}
\caption{Streamwise spectral evolution contours of nonlinear wave packet in the frequency $(f)$ -spanwise wavenumber $(\beta)$ plane (Case~$5$).}
\label{fig:NLFM_FFT}
\end{figure}

The spectral signature of the disturbance evolution (Fig.~\ref{fig:NLFM_FFT}) further highlights the deviations from the linear wave packet.
Stronger planar Mack modes~$(f\sim 82kHz)$ and weaker oblique first modes~$((f,\beta)=(22.6,\pm 10.19))$ are apparent in Fig.~\ref{fig:NLFM_FFT}(a) i.e., the dominant $2D$ high-frequency and oblique lower-frequency waves features observed in the linear case (Fig.~\ref{fig:LFM_FFT}(b), Case~$3$) persist with nonlinear forcing. 
However, the larger amplitudes engender harmonics of the fundamental wave ($f\sim141kHz$) due to the self-sum interaction, $(70.5,0)+(70.5,0)\rightarrow (141,0)$.
The energy is spread into spanwise waves, consistent with the pressure perturbations in Fig.~\ref{fig:NLFM_PP}(d).
Proceeding downstream, the Mack mode and its harmonics are also observed for $x=29$ (Fig.~\ref{fig:NLFM_FFT}(c)), at lower frequencies than at upstream stations.
The presence of low-frequency disturbances indicate laminar mean flow distortion by the wave packet and 
deviations from the linear counterpart (Fig.~\ref{fig:LFM_FFT}(c)) also suggest the onset of secondary instabilities.
Oblique waves at the same frequency as the second mode at $(f,\beta) = 65kHz, \pm 10)$ exhibit large amplitudes.
This further points to fundamental resonance~\cite{herbert1988secondary} as a dominant secondary instability mechanism.
Weak oblique disturbances at half the Mack mode frequency with $(f, \beta)=(32.5kHz,\pm 10)$ are also apparent, and can be interpreted as a signature of subharmonic resonance.
While the above two secondary instabilities are related to the second mode, a third class of first-mode (weak) oblique resonance at $(f, \beta)=(20.6kHz, \pm 7.65)$ is also noted.

\begin{figure}
\centering
\begin{subfigmatrix}{3}
\subfigure[]{{\includegraphics[width=0.33\textwidth]{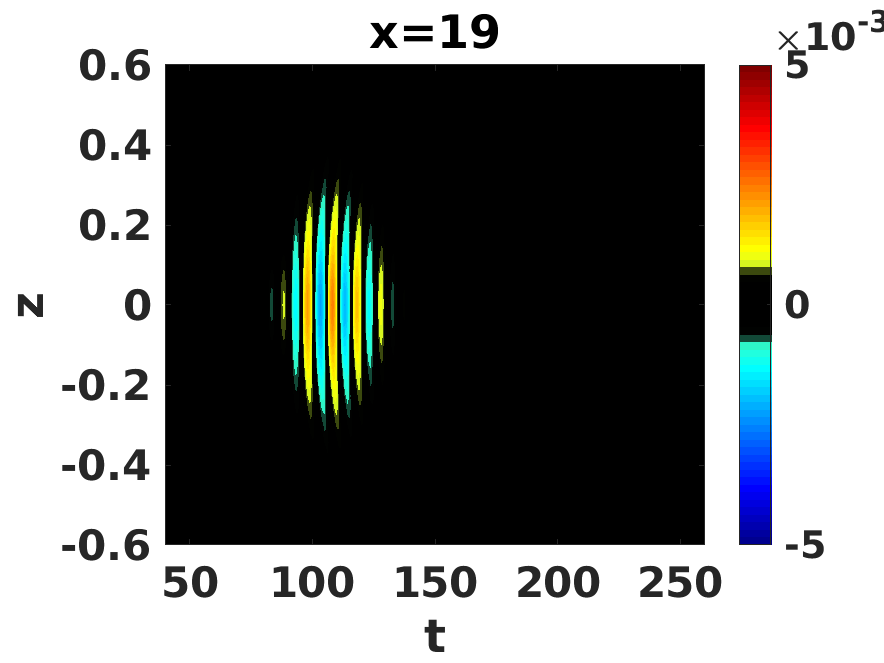}}}
\subfigure[]{{\includegraphics[width=0.33\textwidth]{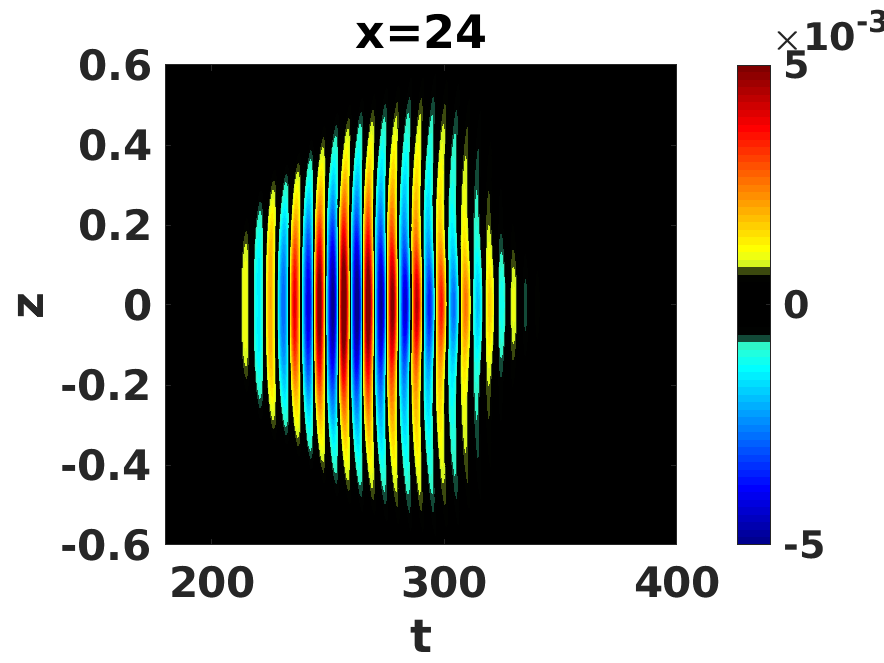}}}
\subfigure[]{{\includegraphics[width=0.33\textwidth]{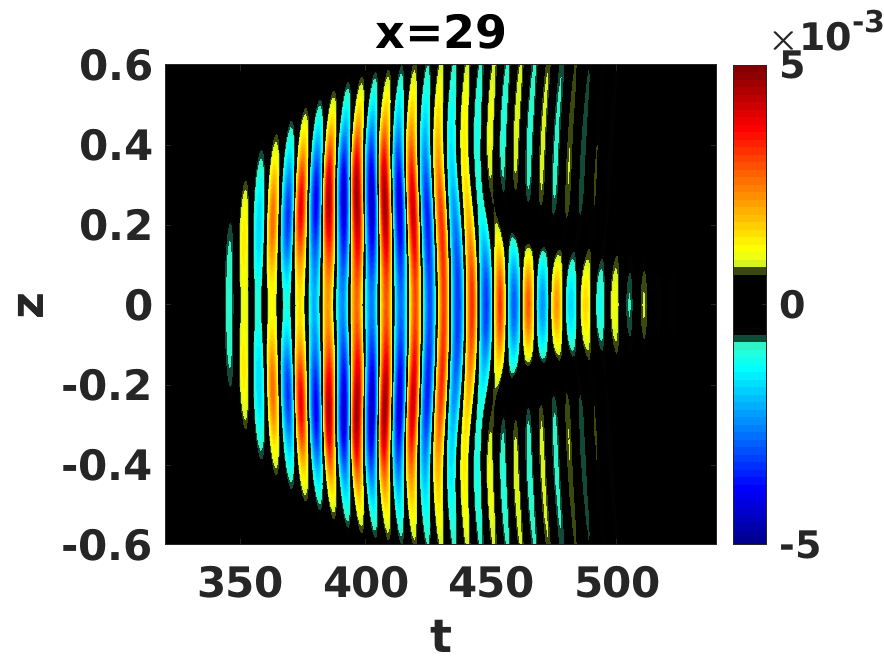}}}
\subfigure[]{{\includegraphics[width=0.33\textwidth]{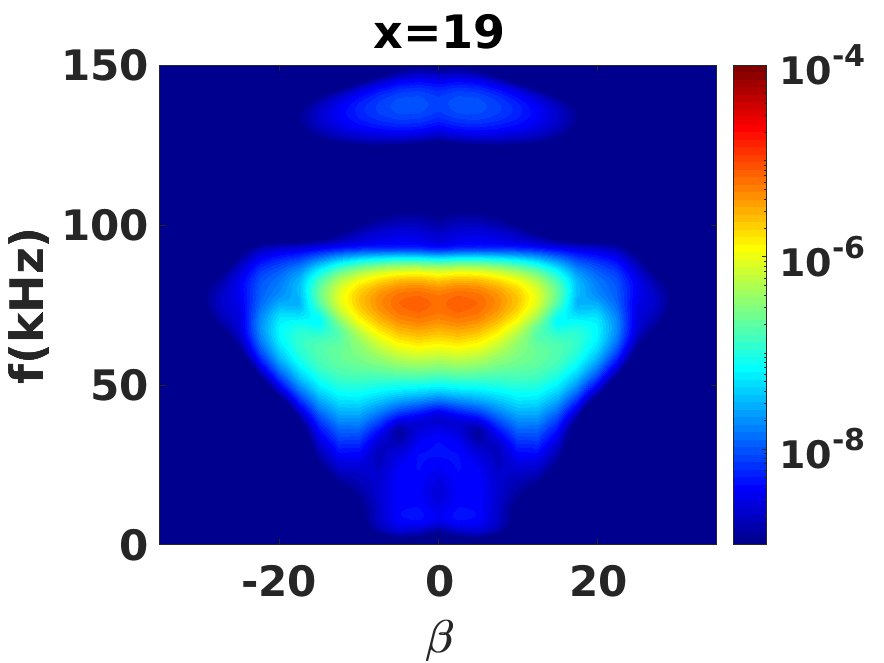}}}
\subfigure[]{{\includegraphics[width=0.33\textwidth]{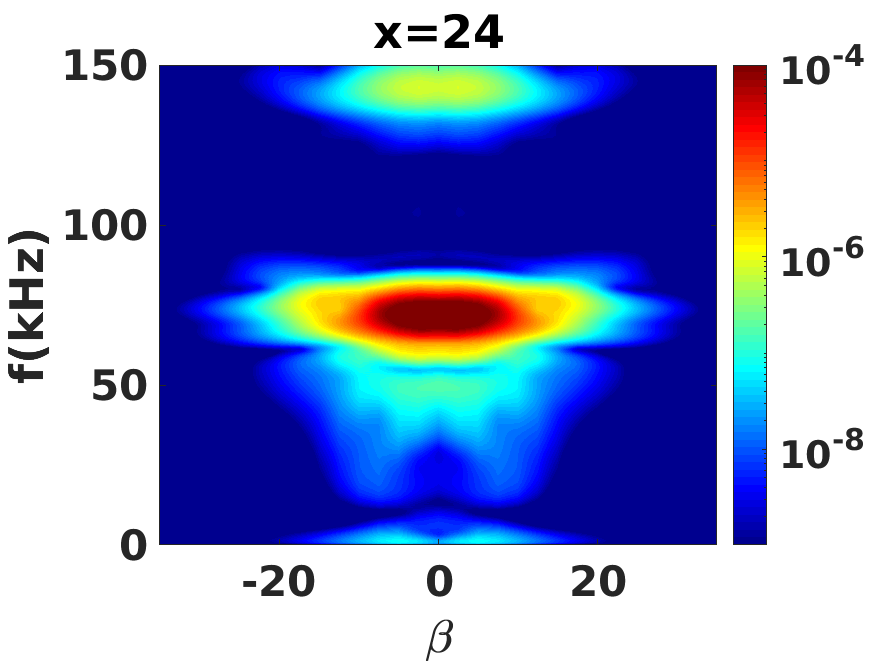}}}
\subfigure[]{{\includegraphics[width=0.33\textwidth]{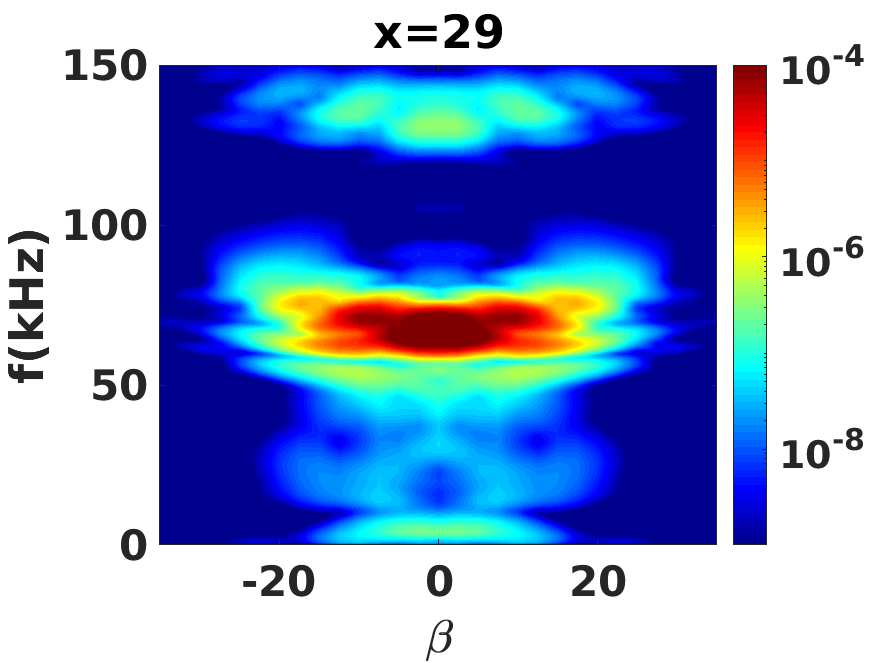}}}
\end{subfigmatrix}
\caption{Nonlinear wave packet evolution for disturbances triggered at $x=16.2$ (Case~$6$).
The wall pressure perturbation contours are shown in $(a,b,c)$.
The corresponding spectral content is plotted in $(d,e,f)$.}
\label{fig:NLSM_PP}
\end{figure}

Case~$6$ examines effect of location of nonlinear wavepacket initiation on perturbation evaluation.
The corresponding wall pressure perturbation and spectral signatures are shown in Fig.~\ref{fig:NLSM_PP}.
An arrowhead-shaped wave packet is again observed in Fig.~\ref{fig:NLSM_PP}(a), similar to its linear counterpart (Fig.~\ref{fig:LSM_PP}(a)) and a weaker oblique wave signature arises in the spectra (Fig.~\ref{fig:NLSM_PP}(d)).
Additionally, higher amplitude forcing also manifests harmonics of the Mack modes.
As it develops downstream (Fig.~\ref{fig:NLSM_PP}(b)), the pressure perturbation signature deviates from the linear wave packet, especially in the observation of a trailing edge protrusion and enhanced spanwise spreading.
This is also reflected in its spectral signature (Fig.~\ref{fig:NLSM_PP}(e)) with oblique waves arising at the second-mode frequency.
Other effects are the emergence of oblique first-mode waves and further lower-frequency disturbances.
This difference from the low-amplitude case may be attributed to higher receptivity coefficients with higher-amplitude forcing.
Hence, first-mode waves could be triggered with a sufficiently strong disturbance source downstream of the ogive-cylinder junction.
At $x=29$ (Fig.~\ref{fig:NLSM_PP}(c)), the three-legged structure of the wave packet is evident, similar to Fig.~\ref{fig:NLFM_PP}(c).
The spectral distribution suggests the dominance of the fundamental resonance with $(f, \beta)=(65.23, \pm 10)$.
The evolution of the nonlinear wave packet with the two forcing locations considered suggests fundamental resonance as the dominant secondary instability mechanism.

\begin{figure}
\centering
\includegraphics[trim=1 70 30 20,clip,width=0.75\textwidth]{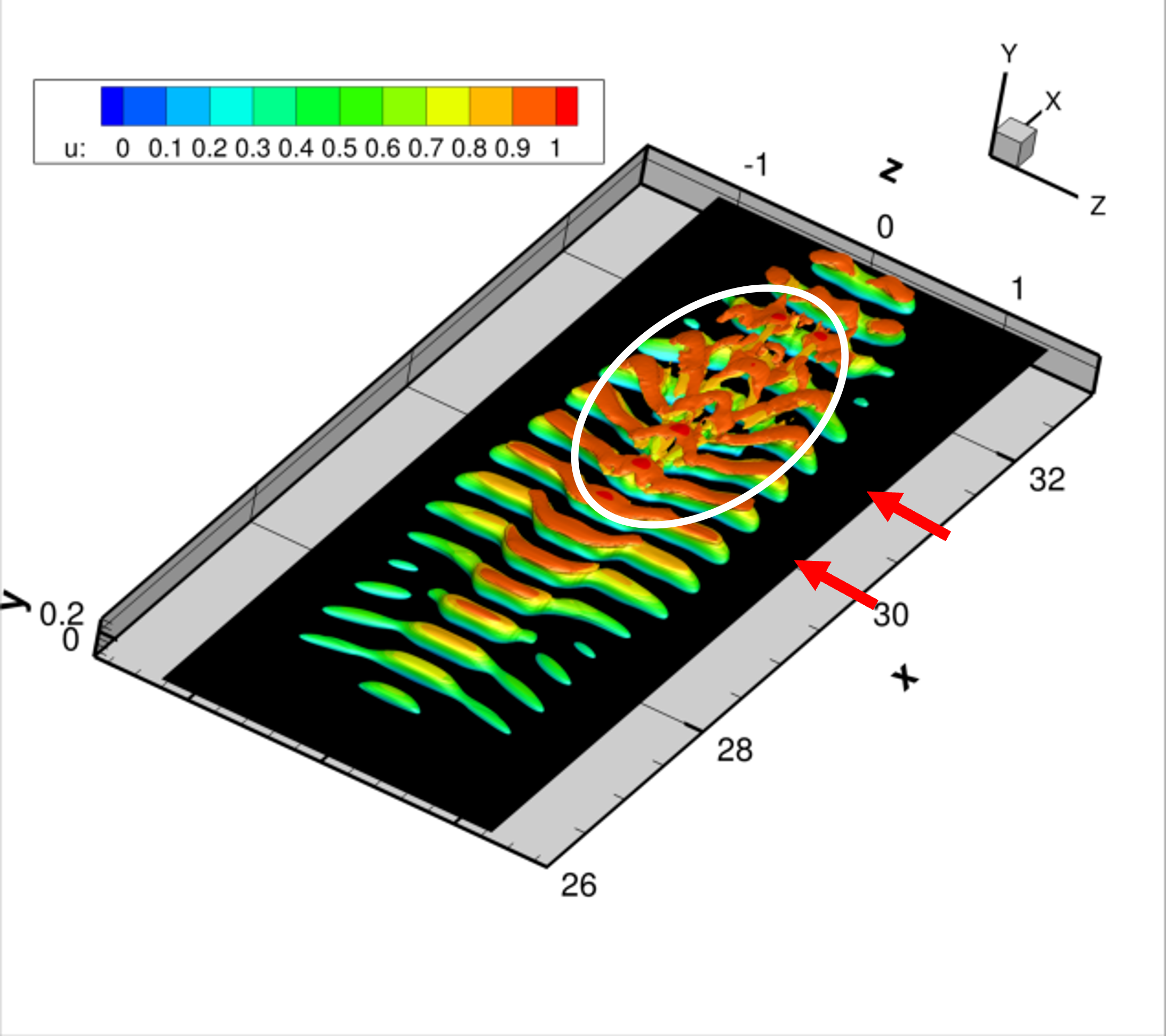}
\caption{Instantaneous vortical structures identified by Q-criterion $(Q=0.2)$ isosurfaces extracted at near the trailing end of the domain for Case~$5$.}
\label{fig:NL_Qcrit}
\end{figure}

The signature of vortical structures in the nonlinear regime for Case~$5$ is shown in Fig.~\ref{fig:NL_Qcrit}.
Similar to the linear wave packet, $2D$ roller structures are observed close to the wall.
These are saturated because of the high forcing amplitudes, and develop a slightly oblique signature in the trailing part.
The primary instabilities (Mack modes) after reaching sufficiently large amplitudes transfer their energies to secondary instabilities, which manifests as lambda vortices at $x=30$ around $z=0$.
%Significant nonlinearities are noticeable at the leading edge of the wave packet, where lambda-type vortices are observed to ride over the near-wall roller structures.
Further downstream $(x=31)$, these amplify by stretching in the streamwise direction along with some spanwise deformation.
These lambda-patterns are (generally) aligned with each other, visually confirming the dominance of fundamental resonance.

\begin{figure}
\centering
{{\includegraphics[trim=0 40 10 60, clip, width=0.75\textwidth]{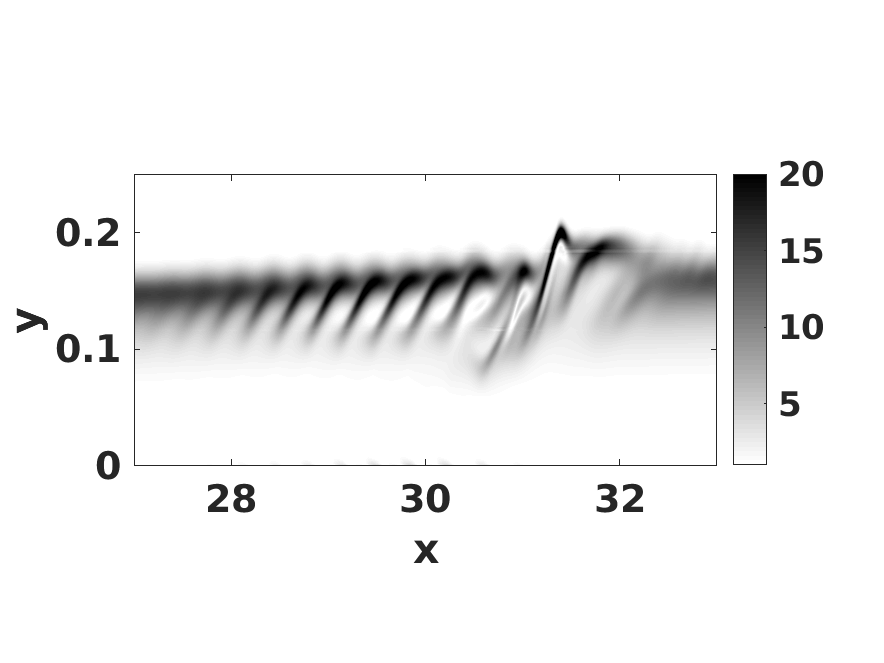}}}
\caption{Pseudo-Schlieren in the symmetry plane of instantaneous vortical structures (Case~$5$).}
\label{fig:NL2D_gradrho}
\end{figure}

The density gradient of the nonlinear wave packet (at $z=0$) plotted in Fig.~\ref{fig:NL2D_gradrho} also shows rope-like structures in $x\in [28,30]$.
These structures extend in the wall normal direction at $x=31$ and resembling late transitional features reported by Hill et al.~\cite{luke2022experimental}.
Their origin could be linked to the strong secondary instabilities observed at this location evident in Fig.~\ref{fig:NL_Qcrit}.

\begin{figure}
\centering
\begin{subfigmatrix}{2}
\subfigure[]{{\includegraphics[width=0.45\textwidth]{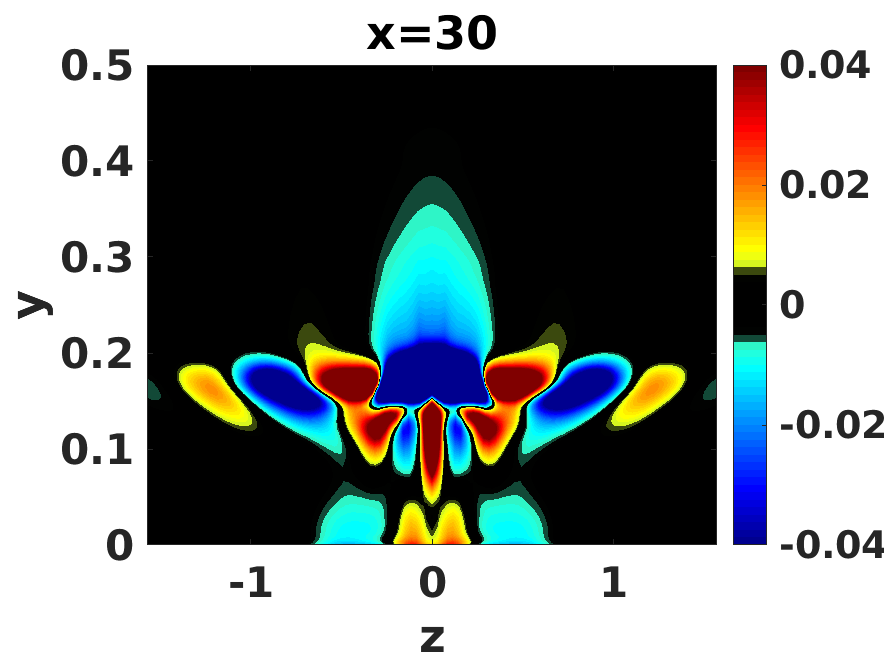}}}
\subfigure[]{{\includegraphics[width=0.45\textwidth]{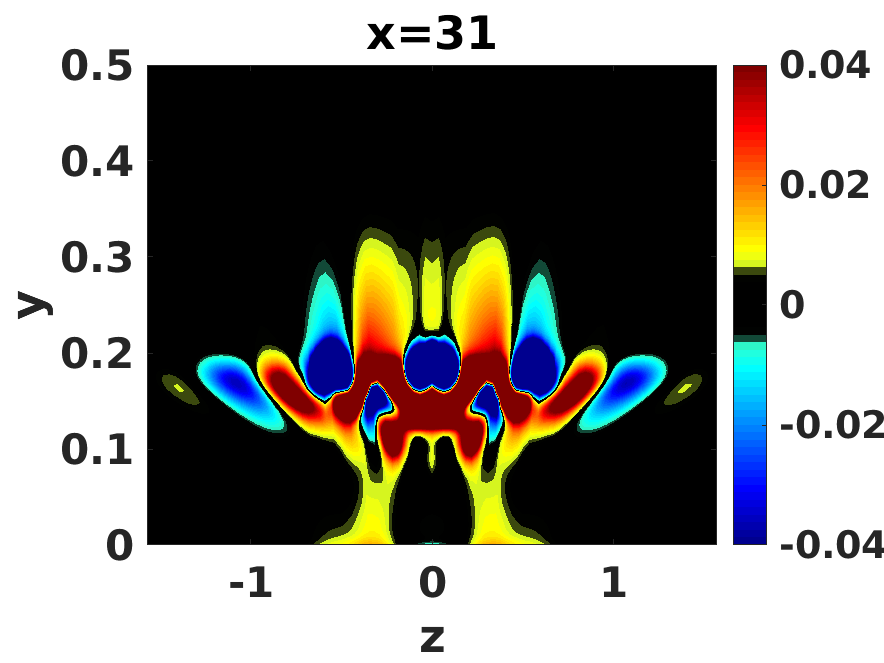}}}
\end{subfigmatrix}
\caption{Density perturbation contours in the cross-flow planes of instantaneous vortical structures (Case~$5$).}
\label{fig:QRX}
\end{figure}

The role of instabilities in generating the Schlieren signature can be explained through the density perturbation contours of the wave packet at various cross-flow planes, as shown in Fig.~\ref{fig:QRX}.  
At $x=30$ (Fig.~\ref{fig:QRX}(a)), the density perturbation is strongest at locations away from the wall, $(y,z)\sim(0.2,0)$ and is surrounded by oblique waves at $(y,z)\sim(0.2,-1)$ similar to the prior observations with the linear wave packet (Fig.~\ref{fig:LinQR}).
New smaller wavelength disturbances are observed in $(y,z)\sim(0.1,0.2)$, suggesting deformation of the Mack mode.
Spanwise spreading of the Mack modes is noticeable closer to the wall at $(y,z)\sim(0,0)$.
In Fig.~\ref{fig:QRX}(b) $(x=31)$, a significant deformation of the Mack mode structure is observed at $(y,z)\sim(0.2,0)$, with a reduction in its spanwise extent and the development of new adjacent lambda-shaped structures.
The wall-normal extension in the pseudo-Schlieren is due to these instabilities and can be considered to be the imprints of an energy transfer between primary and secondary instabilities.
The resulting spanwise modulation signature is observed on the wall for $(y,z)\sim (0,\pm 0.5)$; however the waves at $(y,z)\sim (0.2,-1)$ do not significantly modulate the base flow.
%???do you mean "modulate the mean flow significantly?"
%HG: Both are okay. I modified it to base flow now.

\section{Conclusions}
\label{sec:conclusions}
Linear and nonlinear disturbance evolution over the frustum of ogive-cylinders at Mach~$6.1$ are examined with high-fidelity numerical simulations to augment prior linear stability analyses.
%\todo{Would be useful to say ``to augment prior stability analyses'' implicitly referring to others who may have done this that way. HG:Done}
The geometry and freestream parameters are obtained from recent experiments conducted at the Air Force Research Laboratory Mach~$6$ wind tunnel.
The linear response of the boundary layer to freestream stochastic pressure perturbations indicates that the most amplified frequencies closely match the experimental observations, and the evolution features of the intermittent wave packets are accurately captured.
Planar Mack modes are identified as the dominant primary instabilities.
Oblique first modes are also observed; these amplify over a relatively longer streamwise length but at an order of magnitude lower amplitude than Mack modes.
%However, first-mode waves amplify over a relatively longer streamwise length.

A more controlled scenario is examined by considering wave packet evolution.
For this, a broadband wave packet centered around a second-mode wave frequency is excited at different locations on the ogive-cylinder.
When the wave packet is initiated closer to the ogive-cylinder junction, an arrowhead-shaped wave packet evolves with a structure similar to those observed previously on hypersonic flat plates.
The spectral content of these disturbances identify dominant second-mode and weaker first-mode instabilities.
When the wave packet is introduced at a further downstream location, while the second-mode waves dominate the primary instabilities, the receptivity of the first-mode waves was found to be very low.

The nonlinear modulation of broadband disturbances is then examined through high-amplitude wave packet actuation.
When the source of disturbances closer to the ogive-cylinder junction, the wave packet modulates in the azimuthal direction, forming three-legged structures due to the spanwise curvature effects.
Spectral analysis shows  strong fundamental resonance, followed by weak subharmonic and oblique resonances as secondary instability mechanisms. 
When the nonlinear wave packet is triggered further downstream, both first- and second-mode waves are observed.
The three-legged structure of the wave packet is preserved, and fundamental resonance is the dominant secondary instability mechanism.
Experimentally observed rope-like structures of Mack modes and their wall-normal deformation in late transitional stages are captured in pseudo-Schlieren contours.
These deformations are attributed to the onset of secondary instabilities over Mack modes closer to the edge of the boundary layer.
In future work, the transition to turbulence through fundamental resonance of second modes will be examined to assess the differences from axisymmetric cones. 

\section*{Acknowledgements}
This research was supported by the Office of Naval Research (Grant: N00014-21-1-2408) monitored by Dr. E. Marineau with R. Burnes as the technical point of contact.
Discussions with Dr. M. Choudhari, Dr. P. Paredes, and Prof. C. Hader are gratefully acknowledged.
The opinions, findings, views, conclusions, or recommendations contained herein are those of the authors and should not be interpreted as necessarily representing the official policies or endorsements, either expressed or implied, of ONR or the U.S. Government.
The simulations were carried out using resources provided by the U.S. Department of Defense High Performance Computing Modernization Program, Ohio Supercomputer Center, and FSU Research Computing Center. 
% Several figures were made using FieldView software with licenses obtained from the Intelligent Light University Partnership Program.

\bibliography{bibfile}

\end{document}